\documentclass[]{aa} 
 
\usepackage{graphicx} 

\usepackage[normalem]{ulem} 
\usepackage{txfonts} 
\usepackage{lscape} 
\bibpunct{(}{)}{;}{a}{}{,} 
\DeclareGraphicsExtensions{.ps,.eps,.pdf,.gif,.jpg} 
\begin{document}

   \title{Searching for galaxy clusters in  the Kilo-Degree Survey}

   \author{M. Radovich \inst{1} 
          \and E. Puddu \inst{2} 
          \and F. Bellagamba \inst{3,4} 
          \and M. Roncarelli \inst{3,4} 
                    \and L. Moscardini \inst{3,4,5} 
                    \and S.Bardelli \inst{4} 
          \and A. Grado \inst{2} 
          \and F. Getman \inst{2} 
          \and M. Maturi \inst{6} 
          \and Z. Huang \inst{2} 
          \and N. Napolitano  \inst{2} 
          \and J. McFarland \inst{7} 
          \and E. Valentijn\inst{7}    
          \and M. Bilicki \inst{8}        
          } 
 
\institute{ 
        INAF - Osservatorio Astronomico di Padova, vicolo dell'Osservatorio 5, I-35122 Padova, Italy 
        \and INAF - Osservatorio Astronomico di Capodimonte, Salita Moiariello 16, I-80131 Napoli, Italy        
        \and  Dipartimento di Fisica e Astronomia, Alma Mater Studiorum - Universit\`{a} di Bologna, viale Berti Pichat 6/2, I-40127 Bologna Italy 
        \and INAF - Osservatorio Astronomico di Bologna, via Ranzani 1, I-40127 Bologna, Italy  
        \and INFN - Sezione di Bologna, viale Berti-Pichat 6/2, I-40127 Bologna, Italy  
    \and Zentrum f\"ur Astronomie, Universitat\"at Heidelberg, Philosophenweg 12, D-69120 Heidelberg, Germany 
    \and Kapteyn Astronomical Institute, P.O. Box 800, 9700 AV Groningen, The Netherlands 
    \and Leiden Observatory, Leiden University, P.O. Box 9513 NL-2300 RA Leiden, The Netherlands 
         } 
   \date{Acceptance date: 23/12/2016} 
 
  \abstract 
   {} 
   {In this paper, we present the  tools used to search for galaxy  clusters in the Kilo Degree Survey (KiDS), and our first results. } 
   {The cluster detection is based on an implementation of the optimal filtering technique that enables us to identify clusters as over-densities in the distribution of galaxies using their positions on the sky, magnitudes, and photometric redshifts.  The contamination and completeness of the cluster catalog are derived using mock catalogs based on the data themselves.  The optimal signal to noise threshold for the cluster detection is obtained by randomizing the  galaxy positions and selecting the value that produces a contamination of less than 20\%.  Starting from a subset of clusters detected with high significance at low redshifts, we shift them to higher redshifts to estimate the completeness as a function of redshift: the average completeness is $\sim$ 85\%.  An estimate of the mass of the clusters is derived using the richness as a proxy.} 
   { We obtained  1858 candidate clusters with redshift 0 $< z_c < $ 0.7 and mass $10^{13.5} < M_{500} < 10^{15}$ $M_\odot$ in an area of 114 sq. degrees (KiDS ESO-DR2). A comparison with publicly available Sloan Digital 
Sky Survey (SDSS)-based cluster catalogs  shows that we match more than 50\% of the clusters (77\% in the case of the redMaPPer catalog). We also cross-matched our cluster catalog with the Abell clusters, and clusters found by XMM and in the Planck-SZ survey; however, only a small number of them lie inside  the KiDS area currently available.} 
   {} 
 
 
   \maketitle 
%
 
\section{Introduction} 
\label{sec:1} 
 
Clusters of galaxies are described as the most massive  collapsed structures in  the Universe. They represent a powerful tool for  cosmological studies \citep{Allen2011}, making it possible to probe  
 the formation history of cosmic structures at different redshifts, and to constrain the measurement of cosmological parameters, such as 
the matter density parameter $\Omega_M$ and the power spectrum normalization $\sigma_8$ \citep[see e.g.][and references]{1993Natur.366..429W,1998MNRAS.298.1145E, sartoris2016}.  
In this perspective, it becomes crucial that  photometric surveys be able to supply  
a statistically significant sample for the detection of clusters  over large sky areas, compared to X--rays surveys, for example, that cover smaller patches. 
Among the available surveys of different sizes and depths, one of the landmarks is the Sloan Digital 
Sky Survey (SDSS) \citep[SDSS]{York2000}, probing the low-redshift universe with an imaging sky coverage of $14,555$ sq.  
degrees. Ongoing programs, like the Kilo Degree Survey \footnote{\url{http://kids.strw.leidenuniv.nl}} \citep[KiDS,][]{DeJong2013} and the Dark Energy Survey \citep[DES,][]{DES}, will provide samples of clusters spanning a wider range  of redshift and mass, thanks to their superior depth. 
When completed, KiDS will cover 1500 sq. degrees in the $ugri$ bands, with optimal seeing conditions in the $r-$band ($< 0.8$\arcsec);  
DES will cover a larger area (5000 sq. degrees in $griZY$), with an image quality between SDSS and KiDS, and  typical seeing $\sim 1$\arcsec \citep{Melchior2015}.  Both surveys will make it possible to extend  the search of galaxy clusters to redshifts $z \sim 0.9$ \citep[see][for results based on the DES Science Verification data]{2016ApJS..224....1R}. Finally, the European Space Agency Cosmic Vision mission \textit{Euclid} \citep{Laureijs2011}, planned for launch in 2020, will be able to detect galaxy clusters up to redshift $z=2$ and to calibrate the cluster mass proxy with an accuracy $<$ 10, 30\%, using weak lensing and spectroscopic data, respectively  \citep{sartoris2016}.

In this paper we discuss the first results of the galaxy cluster search 
in the KiDS survey, based on the KiDS ESO-DR2 data release \citep{DeJong2015} (KDR2 hereafter); improvements  due to the availability of larger areas will be discussed in following papers. The cluster search method is based on  \citet[][B11 hereafter]{Bellagamba2011}. Compared to other methods  employing the identification of the red sequence, for example, this approach presents the advantage that it does not search for a specific feature  (e.g., color and brightness) of the cluster member galaxies. Instead, it enables us to simultaneously use the available information on the spatial distribution, magnitudes and photometric redshifts of the galaxies to find over-densities related to galaxy clusters. In addition to cluster searches using only optical data, this approach  has been also used in cluster identifications based on X-ray  \citep[e.g.][]{Pace2008, Tarrio2016} or weak lensing  \citep{Maturi2005} data.

The paper is organized as follows: the main features of KiDS and its data products are summarized in  Sect.~\ref{sec:kids}; details of the cluster finder algorithm are discussed in Sect.~\ref{sec:algorithm}; the contamination and completeness of the cluster catalog are discussed in  Sect.~\ref{sec:tests}; richness and mass of the clusters are derived as explained in Sect.~\ref{sec:properties}; Sect.~\ref{sec:results} shows the properties of the cluster catalog by comparing it with other cluster datasets based on the SDSS. Section~\ref{sec:other}  presents a summary of the  clusters in the Abell, XMM, and Planck-SZ catalogs, which are located in the KDR2 area (see also Appendix~\ref{app:A}). 
Conclusions are given in Sect.~\ref{sec:conclusions}.   
 
\begin{figure*}[pht] 
        \includegraphics[scale=.6]{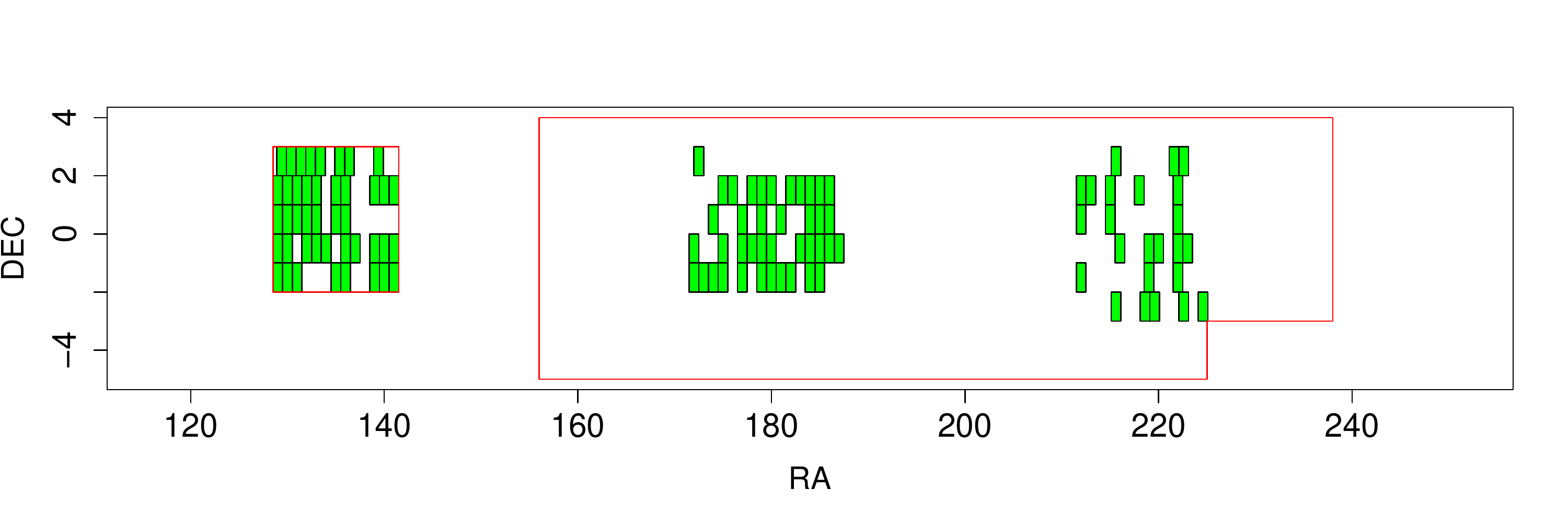}\\ 
        \caption{Position on the sky of the KiDS-N ESO-DR2  tiles (in green). The red boxes show the total KiDS-N planned area.} 
        \label{fig:sky}        
\end{figure*} 
 
\section {The Kilo-Degree Survey} 
\label{sec:kids} 
The Kilo Degree Survey is one of the ESO public surveys being performed with the OmegaCam wide-field camera (1 square degree field of view) 
mounted at the VLT Survey Telescope (VST). KiDS is designed to observe  
an area of 1500 sq. degrees in the {\it ugri} bands, with limiting AB magnitudes at 5$\sigma$ in a 2\arcsec\  aperture of respectively  $24.3$, $25.1$, $24.9$ and $23.8$ mag (KDR2). KiDS is made of two patches, one in the equatorial sky (KiDS-N) and the other around the South Galactic Pole (KiDS-S). 
 
The  data processing and catalog extraction are done by the KiDS consortium using the AstroWISE system \citep{Verdoes2011}.  
An extensive discussion of the survey and reduction techniques are given in KDR2. The  data products  included in the  
public ESO release  are, for each band, the final stacked images, weight maps, and masks flagging regions with known problems (e.g., halos and spikes around bright stars, bad columns, etc.). Catalogs giving source positions and the photometry measured with {\sc SExtractor} are derived both for each band independently and  using the $r-$band as detection image.  The star/galaxy classification is  based on the CLASS\_STAR parameter of {\sc SExtractor} measured on $r$-band images, following the procedure  described in KDR2, Sect.~4.5.1.

Photometric redshifts based on $ugri$ photometry are also available within the KiDS collaboration: they were derived using both template fitting  \citep{Kuijken2015} with the {\sc BPZ} code \citep{Benitez2000}, and a 
machine-learning approach \citep{Cavuoti2015} based on the MLPQNA method. 
The BPZ also provides the full redshift probability distribution function (PDF) that is required in our analysis to properly weight the contribution of galaxies. A discussion on the accuracy of the galaxy redshift distribution produced by BPZ and possible improvements are given by \citet{Choi2015, Hendrik2016}. 
 
The machine-learning approach provides very accurate photometric redshifts (1$\sigma$ uncertainty in $\Delta z/(1+z)$ $< 0.03$), which are less sensitive to uncertainties in photometric zero points, for example. However,  machine-learning photometric redshifts are   reliable  
only in the same parameter space sampled by the spectroscopic training sample, which was based on the SDSS  in  \citet{Cavuoti2015}, and they do not provide  the redshift PDF. 
Work is in progress to address these issues, using a deeper spectroscopic training sample, and developing a novel approach to derive PDFs for machine-learning photometric redshifts \citep{Cavuoti2016}. For this reason, in this analysis we opted for the template fitting photometric redshifts.  We refer to \citet{Kuijken2015} for  details on how  they  were derived: as displayed in their Fig.~12, for $z< 0.7$ the rms scatter in $\Delta z/(1+z)$ is $<0.05$,  and the outlier fraction is $< 10\%$.  At the time the current analysis was done, they were computed only in the KIDS-N tiles  ($\sim$ 114 sq. degrees) overlapping with the  Galaxy and Mass Assembly (GAMA) Survey \citep{Driver2011}.

The tiles available in KDR2 do not cover a contiguous area: for this reason, in this work we analyze each tile independently. Figure~\ref{fig:sky} shows the position on the sky of the tiles used for  this paper. The latest KiDS public release, KiDS ESO-DR3, comprises an area of 440 sq. degrees \citep{Hendrik2016}: the extension of our analysis to the new data, including KIDS-S, is in progress and will be presented in a future paper.   
 
The analysis in this work is based on the sources classified as galaxies in the KiDS catalogs, with an $r-$band magnitude brighter than the limiting magnitude at 10$\sigma$, $m_{10\sigma} \sim 24.2$ mag.  We removed from the catalogs all sources that were detected on spikes and halos nearby bright stars, where the density of spurious detections is higher and would increase the probability of obtaining false positive cluster candidates. To this end, we removed  all detections where one of the following masking flags (see Table 4 in KDR2) is  set to: 1 (readout spike), 2 (saturation core), 4 (diffraction spike), 16 (secondary halo), or 64 (bad pixels). Flagged regions were taken into account in the effective area computation for each KiDS tile.   
  
An initial estimate of the number of expected clusters in KiDS vs. redshift was derived in KDR2 using the  mock catalogs by \citet{Henriques2012} from the Millennium Simulation \citep{Springel2005}. According to this simulation, we would expect to detect $\sim$ 1980 clusters with redshift 0 < z < 0.7 and 13.5 < log($M/M_\odot$) < 15 in an area of 114 sq. degrees ($\sim$ 25900 clusters in the final KiDS area of 1500 sq. deg.). 
 
\section{The cluster finding algorithm} 
\label{sec:algorithm} 
The search for regions with galaxy over-densities tracing clusters  was performed using the Optimal Filtering technique, described in B11.  
A detailed description of the implementation of the algorithm is provided in a separate paper (Bellagamba et al., in preparation). The main idea of this approach is to describe the data in each point of the space as  a sum of a cluster component $M$ and a field component $N$, which acts as noise for the cluster detection. Then, the amplitude $A$ of the cluster component at the point $\vec x_c$ can be optimally estimated from the data $D$ via 
\begin{equation}\label{eq:optimal} 
A(\vec x_c) = \alpha^{-1} {\int \frac {M(\vec x - \vec x_c)} {N(\vec x)} (D(\vec x) - N(\vec x))} d^n x, 
\end{equation} 
where $\alpha$ is a normalisation constant defined as 
\begin{equation} 
\alpha = \int \frac {M^2(\vec x - \vec x_c)} {N(\vec x)} d^n x. 
\end{equation} 
Applying Eq. \ref{eq:optimal} means filtering the data $D$ with a kernel proportional to $M/N$. In our case, the data are:  galaxy positions on the sky, magnitudes in the $r$ band, and photometric redshifts. Thus, we can make the first term of Eq. \ref{eq:optimal}  more explicit as 
\begin{equation} 
A(\vec \theta_c, z_c) \propto \sum_{i=1}^{N_{gal}} \frac {M(\vec \theta_i - \vec \theta_c, m_i) p_i(z_c)}{N(m_i,z_c)}, 
\end{equation} 
where $\vec \theta_c$ and $\vec \theta_i$ are the positions on the sky of the cluster center and of the $i$-th galaxy, respectively,  $z_c$ is the redshift of the cluster, and each galaxy is weighted by its own redshift probability distribution $p_i(z)$. The sum runs virtually over all the $N_{gal}$ galaxies of the catalog. 
By construction, the peaks of $A$ are the positions where the galaxy distribution resembles more the expected one for the cluster and is less likely to be due to  random fluctuations of the background.  
 
In this work, the model $M$ for the cluster is the expected galaxy distribution as a function of radius and magnitude in the $r$ band. This has been constructed from the average properties of clusters with mass $\sim 10^{14} M_\sun$ in the MaxBCG sample, which is derived from SDSS observations \citep{2009ApJ...699.1333H, 2009ApJ...703.2232S, 2009ApJ...702..745H}. The background field distribution is conservatively estimated using the mean density of galaxies as a function of magnitude and redshift in the KiDS data.  
 
For each tile, the amplitude $A$ is measured on a 3D grid which spans $\alpha$, $\delta,$ and $z$ with a resolution of $\sim 250$ kpc spatially and 0.01 in redshift. The resolution in redshift is smaller than the typical uncertainty of the photometric redshifts, in order not to lose any information on the $z$ dimension present in the data. Then, the peaks of this map are detected and their signal-to-noise ratio   (S/N hereafter) is calculated, dividing A by its uncertainty due to the fluctuations in the background and in the cluster galaxy population (see Eq.~9 in B11). In order to avoid multiple detections of the same halo, we build a cylindrical region around each significant peak, following the size-richness relation of \citet{2009ApJ...699.1333H}. All the peaks at lower S/N inside this region are considered `fragments' of the same halo and thus they do not enter the final catalog. 
 
The redshift probability distribution function enables us to weight each galaxy's contribution to the field and cluster components, and assign to each galaxy $i$ the probability $P_{i,j}$ to be a  member of the cluster $j$ as 
\begin{equation} 
P_{i,j} = \frac {A_j M(\vec \theta_i - \vec \theta_j, m_i) p_i(z_j)}{A_j M(\vec \theta_i - \vec \theta_j, m_i) p_i(z_j) + N(m_i,z_j)}. 
\end{equation} 
 
 For each member galaxy, we can finally derive its best-fit template by running BPZ 
again, with the redshift fixed to the cluster redshift. 
 
\section{Contamination and completeness tests} 
\label{sec:tests} 
\subsection{Contamination} 
\label{sec:purity} 
 
\begin{figure}[pht] 
        \includegraphics[width=\columnwidth]{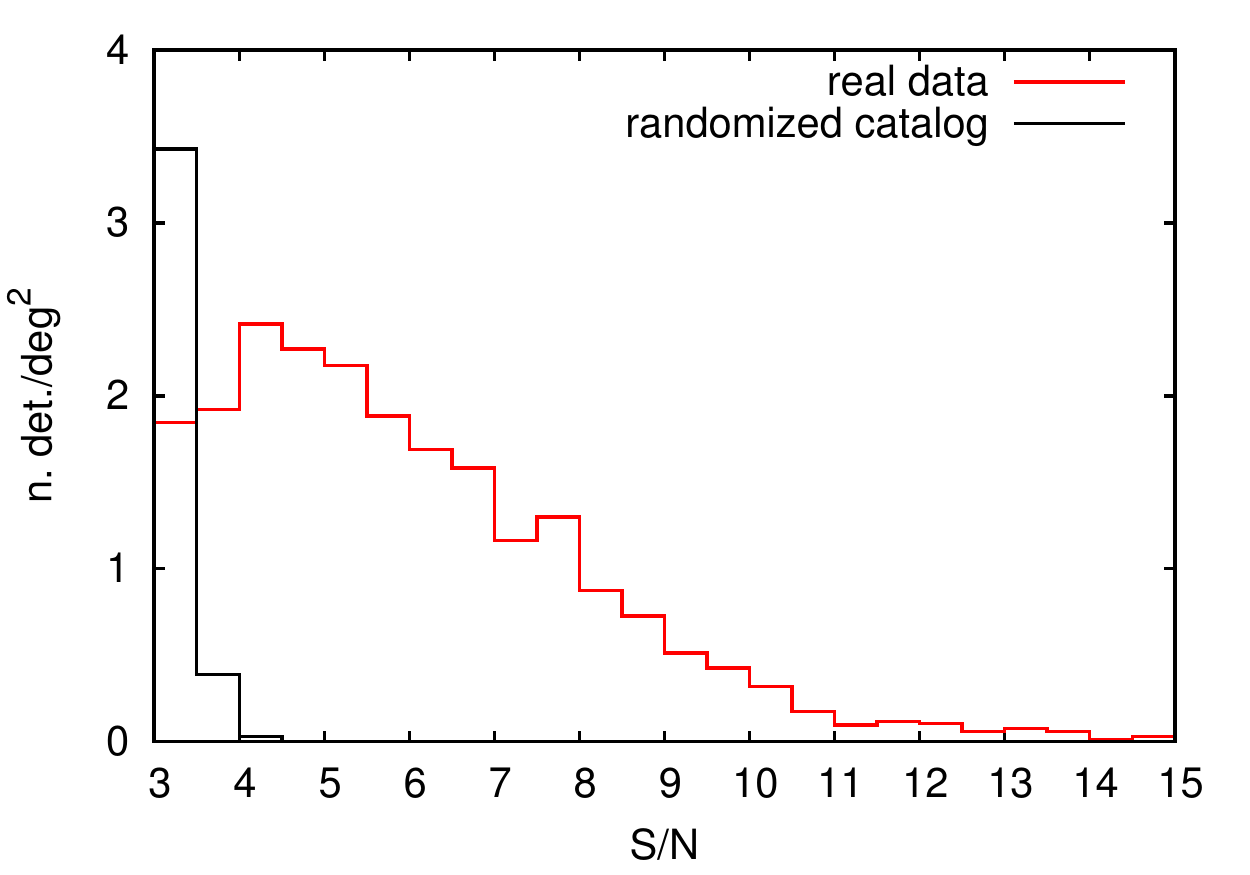}\\       
        \includegraphics[width=\columnwidth]{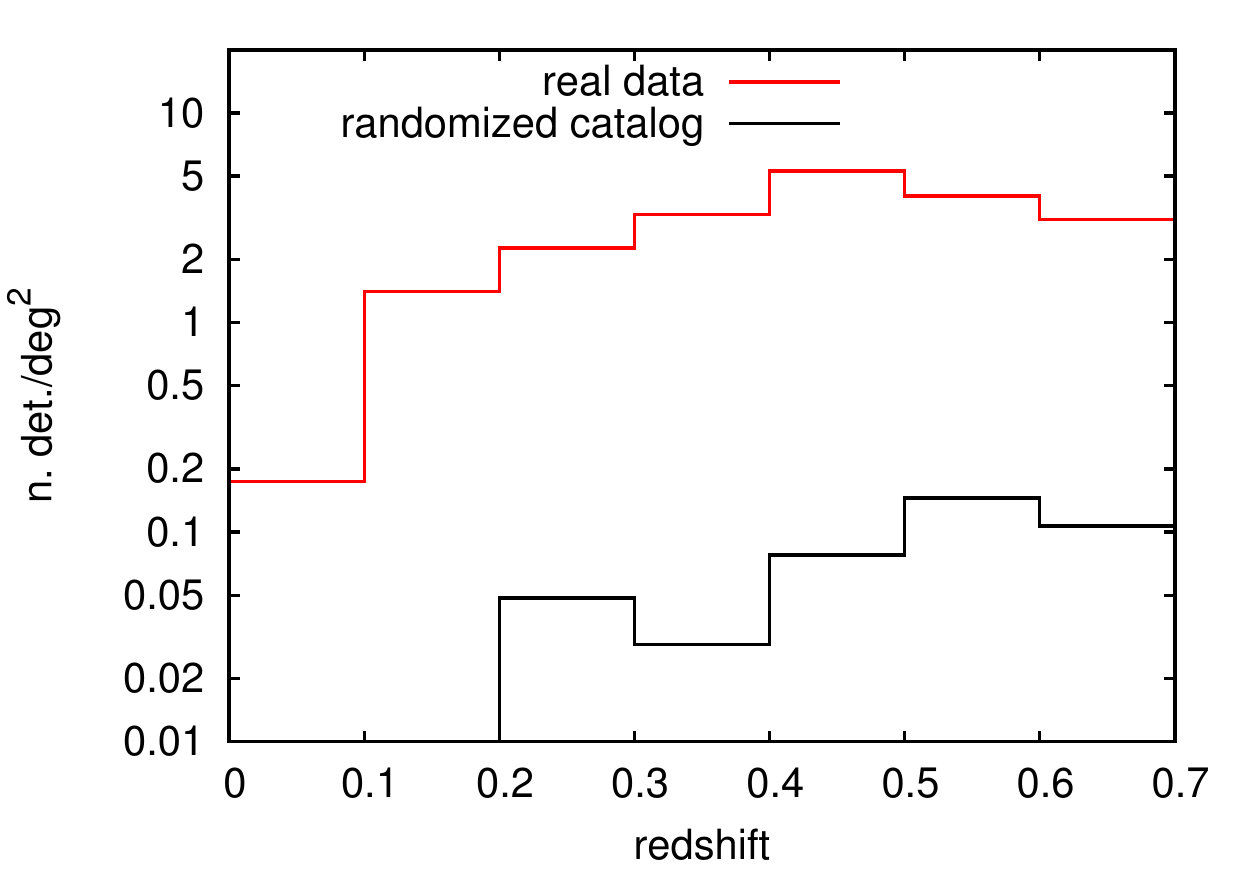}\\ 
        \caption{Number of detections on real data and on randomized catalogs as a function of signal-to-noise ratio (top panel) and redshift (bottom panel). In the bottom panel, only detections with S/N > 3.5 are shown, as this is the limit applied in the final analysis.} 
        \label{fig:purity}        
\end{figure} 
 
The optimal threshold for the S/N should be chosen so that it maximizes the number of true detections ({\em completeness}) and minimizes the number of spurious detections ({\em contamination}). Spurious detections can originate either from galaxies randomly grouped together along the line of sight and mimicking a cluster due to the limited accuracy of photometric redshifts, or from real galaxy associations that are not actually genuine clusters (e.g., groups). In the following, we do not consider any detection associated with physical galaxy groups as spurious, even if its mass is significantly smaller than what is usually considered a cluster. We aim at finding the optimal threshold that minimizes the detection of structures that are produced by random over-densities of objects along redshift and have no physical association. 
 
To this end, we took each tile and  randomized the positions of the galaxies in these regions. In this way, we obtained a catalog that retains all the observational properties of the original dataset (e.g., $p(z)$ distribution, mean density, and luminosity function), but where the structures have been erased. Thus, any possible  detection in such a catalog is by definition a spurious detection. For each randomized catalog, we ran the cluster search as described in the previous section and analyzed the number of detections as a function of S/N. The results are shown in Fig.~\ref{fig:purity}. From the top panel, one can see that in the randomized catalogs there is a relatively large number of detections  produced by chance over-densities of objects, however few of them have S/N > 3.5.  
In real data, there is a high probability that detections at low S/N are considered as fragments of large structures, as described in Sect.~\ref{sec:algorithm} ; therefore, these detections are probably overestimated in the randomized catalogs compared to real data. Nevertheless, based  on this test we selected  S/N = 3.5 as the optimal threshold, giving a contamination of $\sim$ 20\%; with a similar approach, a threshold S/N = 3 was derived in B11.  
In the bottom panel of Fig.~\ref{fig:purity}, we show the redshift distribution of these spurious detections after the cut in S/N is applied.  
We did not consider here the contribution due to correlation from the large scale structure, which may increase the number of spurious detections. 
 
\subsection{Completeness} 
\label{sec:completeness} 
 
In principle, the completeness of the cluster-finding algorithm should be determined using mock galaxy clusters. By analysing the detection output,  one can obtain an estimate of the completeness of the  catalog as a function of redshift, thus providing the characterization of its selection function.  
However, as described in \citet{Ascaso2014}, for example, the semi-analytic galaxy formation  models currently available do not yet fully represent the photometric properties of galaxies in clusters.  
For these reasons, we adopted an alternative approach; we extracted clusters identified at high confidence from the survey itself, shifted them in redshift, and inserted them into the observed galaxy background population to study their detectability. 
 
In detail, we identified two sets of candidate clusters at low redshift ($0.1<z_c<0.3$); a first sample of 19 objects at high  S/N (S/N $>6$) that also match the redMaPPer catalog (see Sect.~\ref{sec:k2rm}), and a second sample of 10 candidate clusters with low S/N (4.8 < S/N< 6). For each cluster, we created its `mock version' by considering its potential cluster members (galaxies with $P_{i,j}>0$, see the definition in Sect.~\ref{sec:algorithm}) and applying a \emph{Monte-Carlo} method to define its actual members; for each $i$-th potential member we independently extracted a random number  $a_i $ between 0 and 1, and only included the galaxy in the mock cluster if $P_{i,j} > a_i$. 
 
In order to model the star-formation history of each mock member, we considered their BPZ library classification, consisting of four CWW templates \citep{coleman80} complemented by two starburst galaxy SEDs computed with GISSEL \citep{bruzual93}. These six templates are well fitted by an exponentially declining star-formation history with an age of 13, 10, 5, 5, 10, and 5 Gyr and $\tau$ (the scale factor of the exponential) of 0.1, 1.0, 3.0, 5.0, 10, and 10, respectively. Each mock galaxy was assumed to have  an age and $\tau$ corresponding to its BPZ template at the redshift of the detection $z_c$;  this allowed us to de-evolve their SEDs using the \cite{bruzual93} recipes, and determine their $r$-band magnitude at the redshift of our choice. We spanned the redshift range $0.2 \leq z \leq 0.75$ with 12 points ($dz=0.05$), thus creating a total of 348  
mock clusters (228 for the high S/N sample and 120 for the low one).  
It is important to  stress that with our approach, the  number of input galaxy members is constant with redshift, while the number of detected galaxy members  varies according to the magnitude evolution. 
 
Then, after choosing a random location in the tile as the mock cluster center, we removed  
the galaxies that fell below the magnitude limit of the tile, and placed the remaining members conserving the cluster geometry, that is, its physical size, and correcting  the angular relative distances accordingly. Each cluster was also randomly rotated around its center. We choose the same tile of the original cluster detection to ensure that the galaxy selection function is  the same. We avoid  putting the cluster center too close to the map border (10  
arcmin) and to the original cluster center (12 arcmin). 
 
As a last step, we need to assign a redshift probability distribution $p(z)$ to each galaxy. This property must be coherent to i) the `true' redshift $z_{\rm gal}$ at  
which we are placing the galaxy, ii) its magnitude $r_{\rm gal}$, iii) its BPZ template  
and iv) the properties of the $p(z)$ of similar galaxies in the KiDS data.  
For this purpose, we need a catalog of galaxies with known spectroscopic redshifts, covering the same redshift and magnitude range as KiDS images. To achieve this, we used data from  the VST-SUDARE/VOICE survey \citep{DeCicco2015} covering the COSMOS field. This survey is deeper than KiDS, therefore we selected a subset of images so that the final depths are the same as in KiDS. The images were processed in the AstroWISE system; $ugri$-band magnitudes, photometric redshifts and best-fit templates were obtained in the same way as for the KiDS data. Catalogs were then matched with the zCOSMOS-bright catalog\footnote{\url{http://www.eso.org/sci/observing/phase3/data_releases/zcosmos_dr3_b2.pdf}} DR3,  providing spectroscopic redshifts for $\sim$20,000 galaxies ($i_{\rm AB} < 22.5$ mag) in the COSMOS field. We thus obtained a catalog of $\sim$7,000 spectroscopic galaxies with redshift $0 < z < 1$. 
 
For any given $z_{\rm gal}$, we  extracted a subsample of spectroscopic galaxies with the same BPZ  
template, and whose spectroscopic redshift $z_{\rm spec}$ satisfied the relation  
$|z_{\rm spec}-z_{\rm gal}|<0.01$. When only a few spectroscopic galaxies satisfied this criterion,  
we increased the tolerance until we reached at least 20 of them: this happened only for some  
templates at high redshift, but we never had to increase the tolerance beyond 0.1. Once this  
subsample was defined, we randomly picked one galaxy with $|r-r_{\rm gal}|<0.1$ and assigned  
its $p(z)$ to the mock galaxy. When no galaxy satisfied the latter criterion, we chose the one with the closest magnitude. This procedure ensures that we respect the   items i)-iv) stated above. 
 
\begin{figure}[pht] 
        \includegraphics[scale=.5]{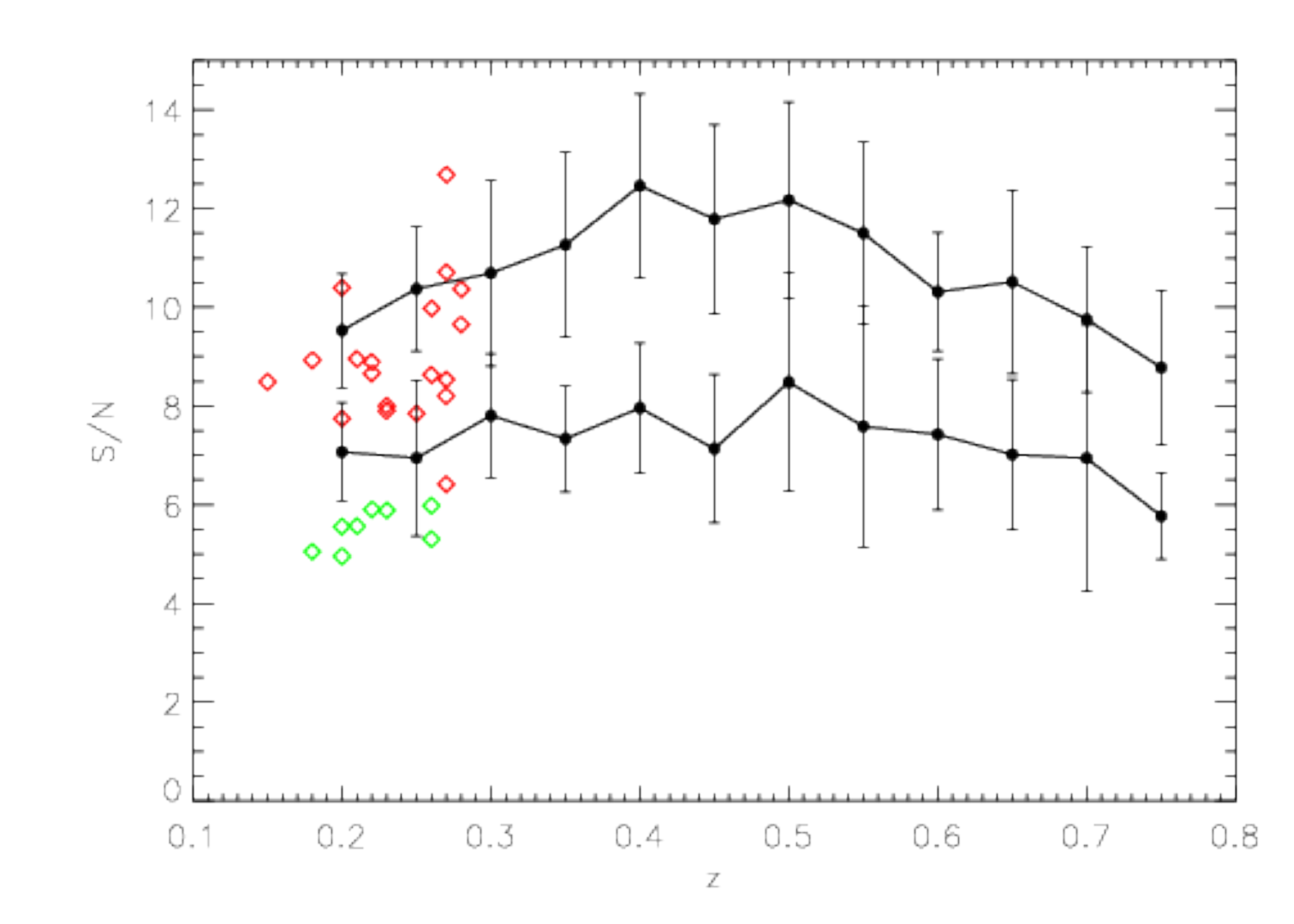}\\ 
        \caption{Average signal-to-noise ratio for our mock clusters as a function of redshift,  
    computed separately for the sample of halos originally detected in the real KiDS data with S/N >6 and S/N < 6 (red and  
    green points, respectively). For each sample, black lines show the average S/N of the detected  
    mock halos at a given redshift, while error bars indicate the r.m.s.} 
        \label{fig:sn_z}        
\end{figure}

 Once this was done, we  ran our algorithm again and checked if we detected the mock  
clusters and, in such a case, the S/N of the detection. The results are shown in  
Fig.~\ref{fig:sn_z}. Only a small number of simulated clusters (33 over 228) were missed by our detection algorithm for the high S/N sample. This suggests a completeness of approximately 85\% over the whole redshift range. When considering the lower S/N sample, the completeness decreases to 70\%. We verified that these non-detections are due to  
the presence of higher S/N detections nearby that masked them out (see the details in  
Sect.~\ref{sec:algorithm}). On the other hand, when the detections are present, the high S/N sample shows almost no evolution of the S/N with redshift, with an increase from $z=0.2$ to $z=0.45$ by approximately 20 per cent and a decrease by the same  
amount down to $z=0.75$. This is the result of a combination of several factors, that include geometry and the difference between the redshift evolution of field and cluster galaxies. On the other hand, no significant trend with redshift is found for the low S/N sample. For this sample, we 
can also observe that a small bias in the determination of the S/N of our mock clusters is present that causes their average S/N to be slightly higher than the original one at $z\sim 0.2$, obtained in real data. This is due to the combination of two effects. First, when computing the average, we consider only mock clusters that have actually been detected, thus with   a higher chance of being at higher S/N. In addition, we verified that our subsample of galaxies with spectroscopic redshift have photometric redshift measurement slightly  
more precise than the average KiDS data at faint magnitudes: this likely induces a small increase   
of the S/N of our mock haloes.  
 
 To summarize, our mock clusters do not change their S/N 
 significantly in the considered redshift interval and therefore we do not expect  
changes in the completeness  up to $z \simeq 0.75$. 
 
\section{Membership, richness, and mass} 
\label{sec:properties} 
 
Cluster members  were selected as those galaxies with $P_{i,j} > 0.2$: this threshold excludes those galaxies that are too faint ($r>24$ mag) and/or too distant from the cluster center ($D > 5$ Mpc).  The brightest galaxy (BCG) is defined as  the one located within 0.5 Mpc of the cluster center derived by the cluster finding procedure.



\begin{figure}[pht] 
        \includegraphics[scale=.5]{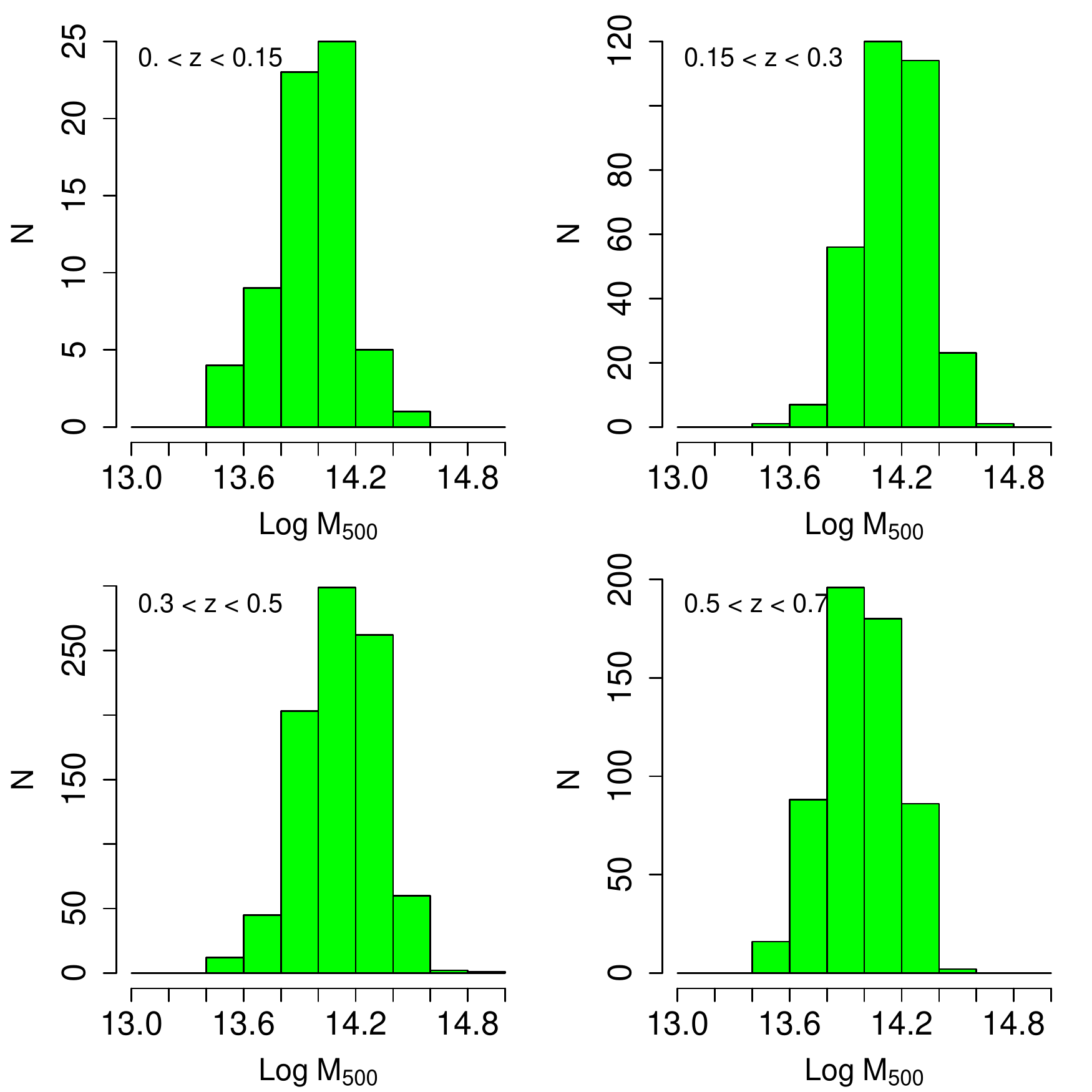}\\ 
        \caption{Distribution of $M_{500}$  derived for the KiDS clusters in different redshift bins.   } 
        \label{fig:m500dist}        
\end{figure}

It is well known that the mass of clusters is well correlated with their total luminosity and richness \citep{Koester2007, Andreon2010}. 
\citet{Andreon2015} showed that the richness ($N$) provides a good  and reliable  proxy to measure the mass ($M$): 
 
\begin{equation} 
\log{M_\Delta} = \alpha + \beta \left( \log{N_\Delta} -2.0 \right) + \gamma \frac{1+z}{1.15}, \label{eq_a15} 
\end{equation} 
where $\Delta$ is the ratio between the cluster average mass density within a radius $R_\Delta$ ,  $\rho_c$ is the critical density of the Universe at that redshift (e.g., $\Delta=$ 200 or 500) and  $N_\Delta$ is the richness derived as outlined below. The term describing the redshift dependence is small ($\gamma \sim -0.1$). 
In their analysis, \citet{Andreon2015} adopted the iterative approach proposed by \citet{Kravtsov2006}. An initial value of $R_\Delta$ was chosen (e.g.,  $R_\Delta = 1 h^{-1}$ Mpc), the richness $N_\Delta\equiv N(< R_\Delta)$ was computed and hence the mass from Eq.~\ref{eq_a15}, assumed to be scatterless.  Then, a new value of $R_\Delta$ was derived, according to: 
        \begin{equation} 
        M_\Delta = \frac{4 \pi}{3} R^3_\Delta \times \Delta \times \rho_c. \label{eq:mass} 
        \end{equation} 
The procedure was repeated until convergence. In \citet{Andreon2015}, the calibration of Eq.~\ref{eq_a15} was based on an X--ray selected sample of 39 clusters with  masses derived by the caustics technique \citep{Rines2013}. An extended catalog of 275 clusters with richness-based masses  was presented in \citet{Andreon2016}. Since these clusters are not covered by the KDR2 tiles and no similar sample is yet available for the current KiDS area, we proceeded as follows. We started from the cluster catalog published by \citet{Wen2015} (WHL15 hereafter), that provides $R_{500}$, $R_{200}$  for 132,684 clusters, using a richness proxy calibrated from  a sample of 1191 clusters with masses estimated by X--ray or Sunyaev-Zeldovich measurements.  By comparing the positions of these clusters with the final area available when  KiDS will be completed, it turned out that  77 of these 1191 clusters will be detectable;  presently however, only three of them fall into the KDR2 tiles, a number too low to produce a reliable fit.    
We therefore used the mass values derived for 230  clusters in the `full' WHL15 catalog  for the calibration, for which $N_{\rm500} > 10$, $z_{\rm WHL15} - z_{\rm KiDS} < 0.05$.  
In this first step, $n_\Delta$ was defined as the number of KiDS cluster members located within a distance from the cluster center $R < R_{\Delta, WHL15}$ and $r-$band absolute magnitudes $M_{\rm abs} + 1.16 z < -20.5$ mag, as in WHL15. $M_{200}$ and $M_{500}$ were derived from Eq.\ref{eq:mass} with the values of  $R_{200}$ and $R_{500}$ in the WHL15 catalog. 
The fit of Eq.~\ref{eq_a15} produced  best-fit coefficients of: $\alpha=14.79 \pm 0.04$, $\beta=0.7 \pm 0.1$ ($\Delta=500$); $\alpha=14.81 \pm 0.03$, $\beta=0.8 \pm 0.1$ ($\Delta=200$);  we did not consider the small redshift--dependent term ($\gamma=0$). 
Using these coefficients, we finally applied the iterative approach outlined before, and obtained $R_\Delta$, $N_\Delta$, and $M_\Delta$ for all KiDS clusters. We did not correct our mass estimate for selection effects such as the Eddington bias \citep[see e.g.,][]{2011PhRvD..83b3015M, 2015MNRAS.450.3633S}: a more detailed discussion on the usage of our catalog for cosmological studies is deferred to the next paper, where more accurate mass estimates based on a larger area  will be available.

\section {Results} 
\label{sec:results} 
 
 We applied the cluster search algorithm to the KDR2  tiles in the KiDS-N area. This gives an effective area of 114 sq. degrees, where we detected 1858 clusters with $0 < z_c < 0.7$, S/N $>$ 3.5, and a mass $M_{500}$, derived as outlined in Sec.~\ref{sec:properties}, between $10^{13.5}$ and $\sim 7 \times10^{14}$ $M_\odot$ (Fig.\ref{fig:m500dist}). 
 
For each cluster we also computed the fraction $f$ of area lost due to masking or to the cluster proximity to the tile borders. The effective area $A_c(R)$ within a given radius is thus:  $A_c(R) = f \pi R^2$. The richness values are corrected by this factor. We selected a subset of 1543 clusters with $f > 0.9$ for which we provide a catalog\footnote{The catalog is available  at \url{http://kids.strw.leidenuniv.nl/DR2}} with the following quantities (Table~\ref{tab:xclustcat}): the  cluster center given by the filter  (\boldmath$\theta$\unboldmath$_c$); the redshift ($z_c$); the signal to noise ratio of the detection (S/N);  the magnitude ($r_{\rm BCG}$) and position (\boldmath$\theta$\unboldmath$_{\rm BCG}$) of the brightest galaxy (the latter is hereafter assumed as the cluster center);  the   $R_{500}$ and $R_{200}$ radii,  the $N_{500}$ and $N_{200}$ richness, and the  $M_{500}$ and $M_{200}$ mass. 

\begin{figure}[pht] 
        \includegraphics[scale=.35]{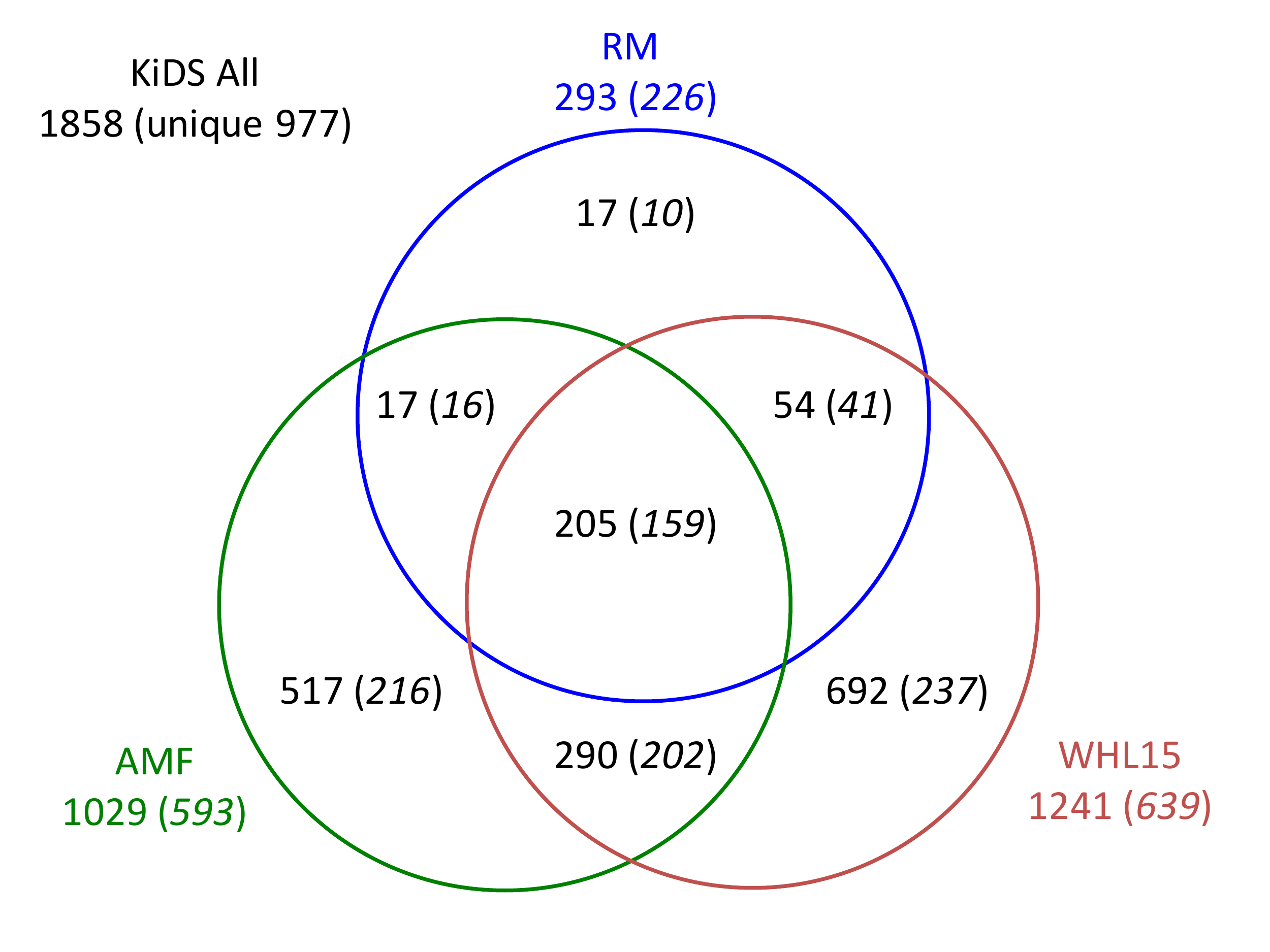}\\ 
        \caption{Venn diagram showing the number of clusters matched among the SDSS-based catalogs  in the KDR2 footprint; the numbers of clusters also matched in KiDS are displayed within brackets.} 
        \label{fig:venn}        
\end{figure} 
 
 \begin{figure}[pht] 
        \includegraphics[scale=.5]{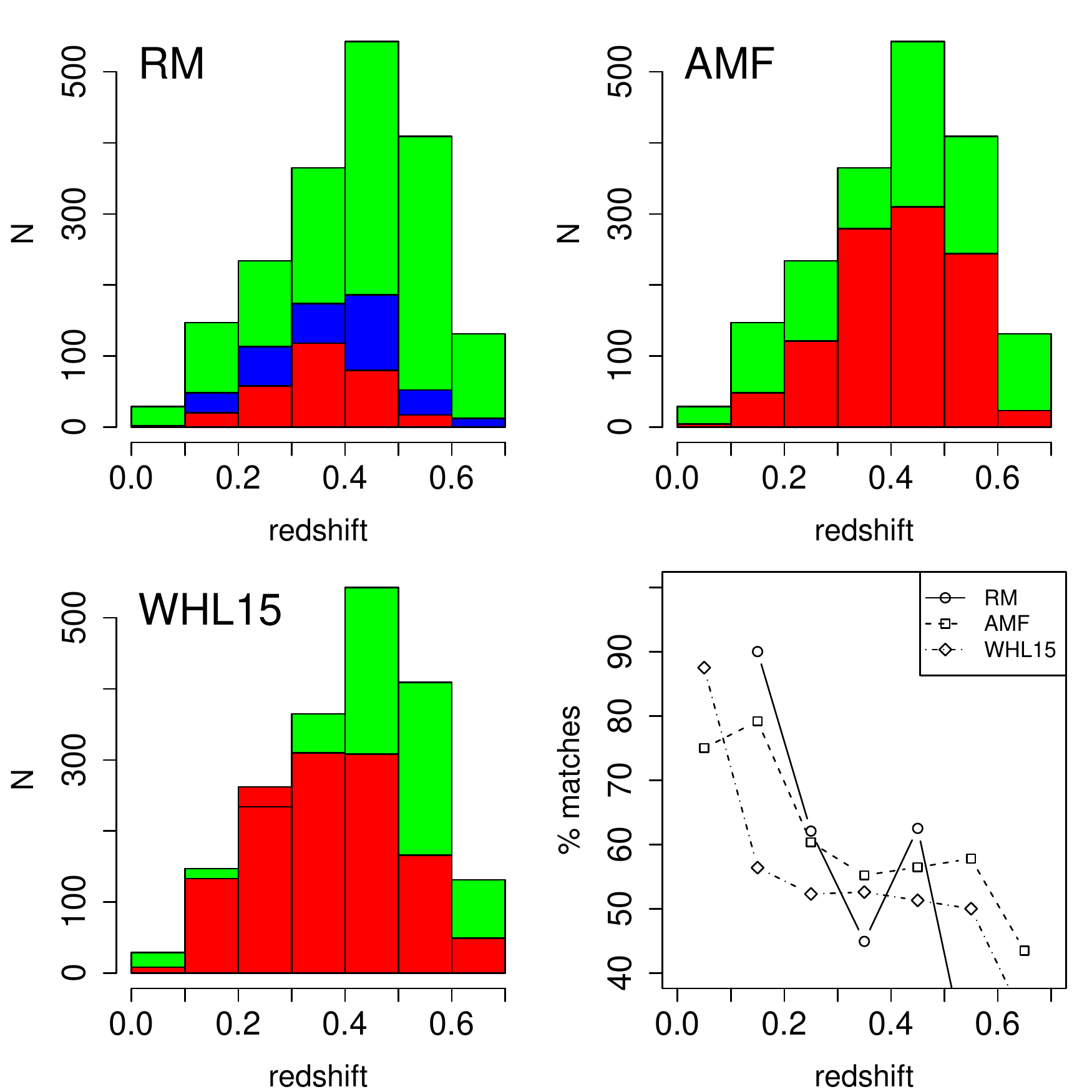}\\ 
        \caption{The first three panels display the redshift distribution of all cluster candidates in the RM, WHL15 and AMF cluster catalogs within the KiDS tiles (red), compared with the KiDS clusters (green). In the first panel (RM), also displayed are the KiDS clusters with a number of early-type galaxies $n_{\rm et}>$ 20 (blue). 
                The last panel shows the fraction of matches only as a function of redshift.} 
        \label{fig:rm_redshift_all}        
 \end{figure} 
  
 \begin{figure}[pht] 
        \includegraphics[scale=.5]{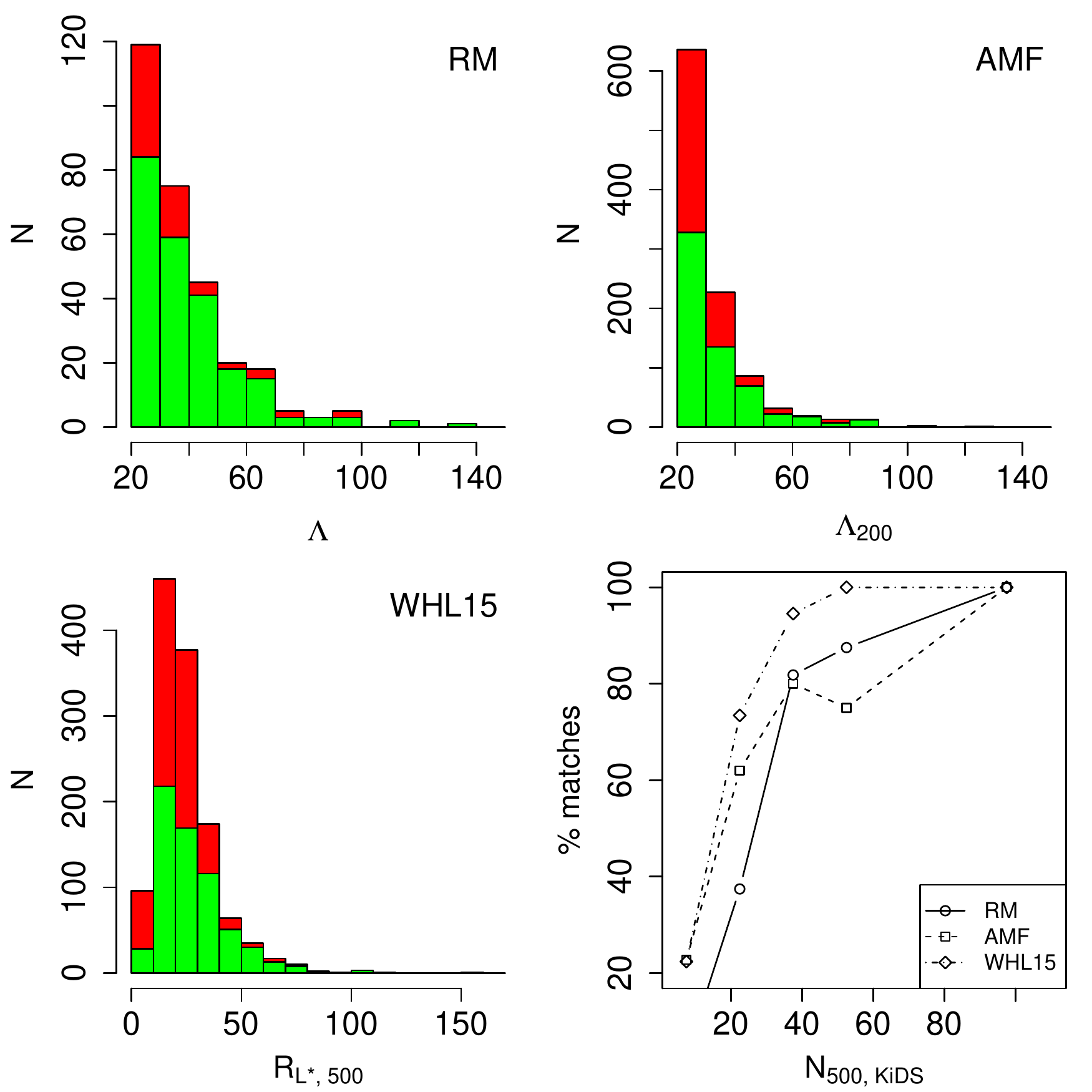}\\ 
        \caption{Distribution of richness estimates for cluster candidates in  
                the RM ($\Lambda$), AMF ($\Lambda_{200}$) and WHL15 ($R_{L*, 500}$) catalogs (in red), with overlaid those matched by KiDS clusters (in green).  The last panel shows the fraction of matches as a function of the KiDS $N_{500}$ richness.} 
        \label{fig:rm_richness_all}        
 \end{figure}

\subsection {Comparison with SDSS cluster catalogs}  
 
In this section, we  compare the KiDS clusters  with those detected in the SDSS in the same area. We use three cluster catalogs derived using different cluster finder algorithms:  
\begin{enumerate} 
\item the redMaPPer catalog \citep[][\ RM hereafter]{Rykoff2014}, based on a red-sequence cluster finder; 
\item the AMF catalog \citep{Szabo2011}, where clusters are identified by an adaptive matched filtering technique  \citep{Dong2008}, similar to what was discussed in  Sect.~\ref{sec:algorithm}; 
\item the WHL catalog first introduced in \cite{Wen2009} and updated in \citet{Wen2012}, \citet[WHL15]{Wen2015}, where clusters are selected using a friend-of-friends algorithm in the (RA, DEC, photo-$z$) space. 
\end{enumerate} 
 
 The number of  clusters detected by each method  depends strongly on the underlying assumptions, as well as on the adopted criteria (e.g., definition of membership, lower limit on richness, center definition). However, while a comparison does not enable us to assess the purity or completeness of a given algorithm, it is helpful to check the consistency of the physical parameters that are derived (redshift, mass, radius). 
 
As a first step, we selected the clusters that fall within the KiDS tiles from the above SDSS-based catalogs. They were then matched to our catalog,  pairing those clusters for which \citet{Szabo2011} centers are closer than 1 $h^{-1}$ Mpc and the difference in redshift is $\Delta z/(1+z) \le 0.1$, considering that the KiDS photo-z rms scatter is $\sim 0.05$ (see Sect.~\ref{sec:kids}). The match was done in such a  way that each cluster from one catalog was matched to only one cluster from the other catalog; in case of multiple matches, the nearest neighbour in both redshift and position is chosen.  Table~\ref{tab:Nsdss} and Fig.\ref{fig:venn} summarize the number of clusters  found in the RM, AMF, and WHL15 catalogs ($N_{\rm SDSS}$), and of those matched  by KiDS clusters ($N_m$); the last  column gives the fraction of matched clusters with separation below $10\arcsec$. The redshift distribution of KiDS and SDSS clusters is displayed in Fig.~\ref{fig:rm_redshift_all}, which also shows the fraction of matched clusters in different redshift bins.  
 
The unmatched clusters can be either real clusters not recovered by our algorithm or spurious detections in the SDSS-based catalogs, or due to an incorrect redshift estimate. An additional issue arises due to the way in which each algorithm handles nearby clusters, merging them in one cluster, or keeping separate substructures.  For instance, \citet{Szabo2011} conclude that, compared to their AMF catalog, the WHL algorithm presented in \citet{Wen2009} produces a fragmentation of the largest clusters into several small clusters.    
Figure~\ref{fig:rm_richness_all} compares the number of matched and unmatched clusters vs. the richness defined in the RM, AMF, and WHL15 catalogs, and  the fraction of matched clusters as a function of the KiDS richness ($N_{500}$). In general, the matching fraction decreases with increasing redshift ($z > 0.3$, Fig.~\ref{fig:rm_redshift_all}) and decreasing richness ($>$70\% for $N_{500}>  40$, $<$ 40\% for $N_{500}< 30$, Fig.\ref{fig:rm_richness_all}), where defining a cluster is more difficult. A similar result is found by  \citet{Szabo2011}, when comparing the AMF and WHL \citep{Wen2009}  catalogs.

\begin{figure}[pht] 
        \includegraphics[scale=.5]{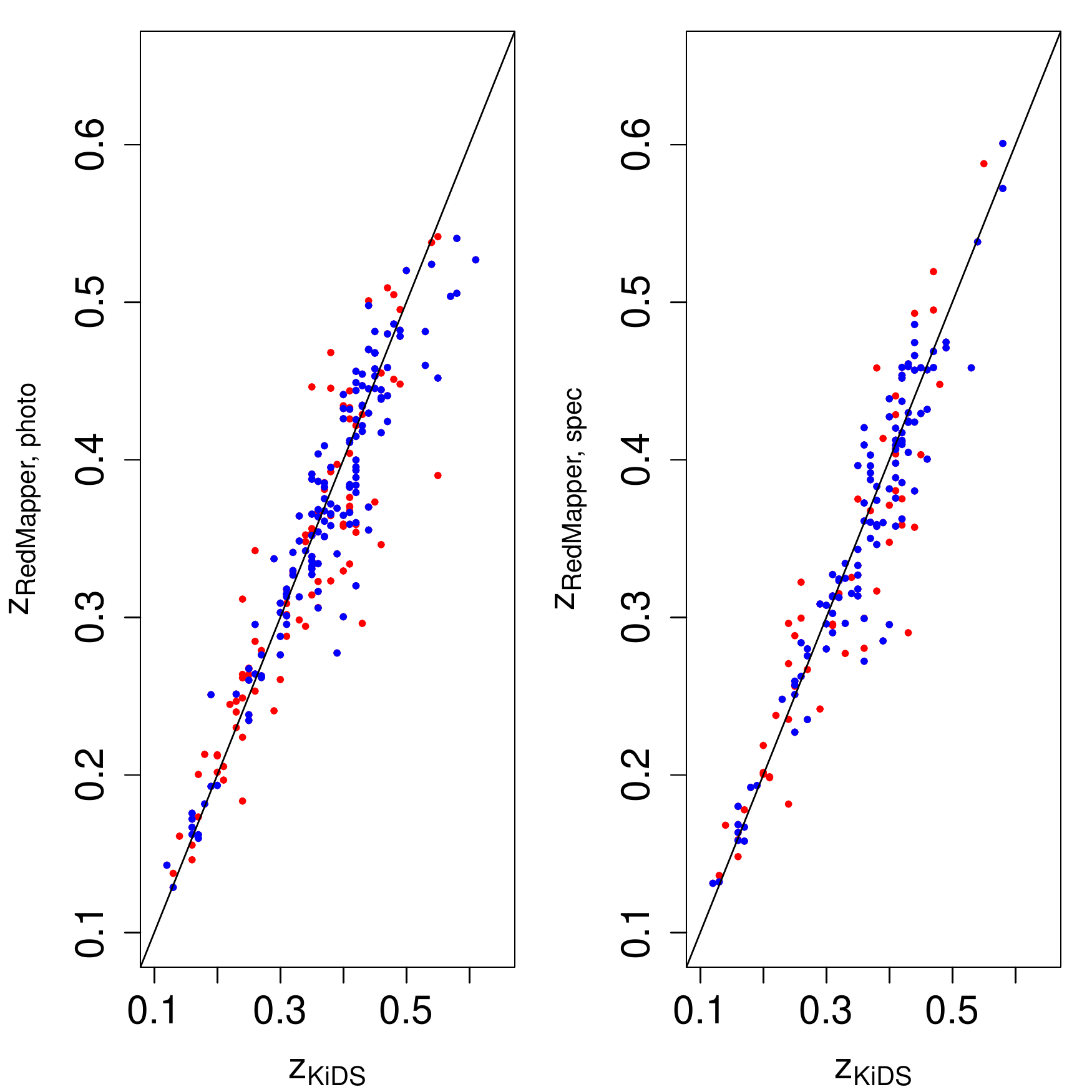}\\ 
        \caption{Comparison of KiDS cluster redshifts with  RM photometric ({\em left}) and spectroscopic  ({\em right}) redshifts.  
                Red dots are clusters matched within 3 arcmin and clusters matched within 1 arcmin are displayed in blue.} 
        \label{fig:rm_zcomp}        
        \end{figure} 
         
\begin{figure}[pht] 
        \includegraphics[scale=.5]{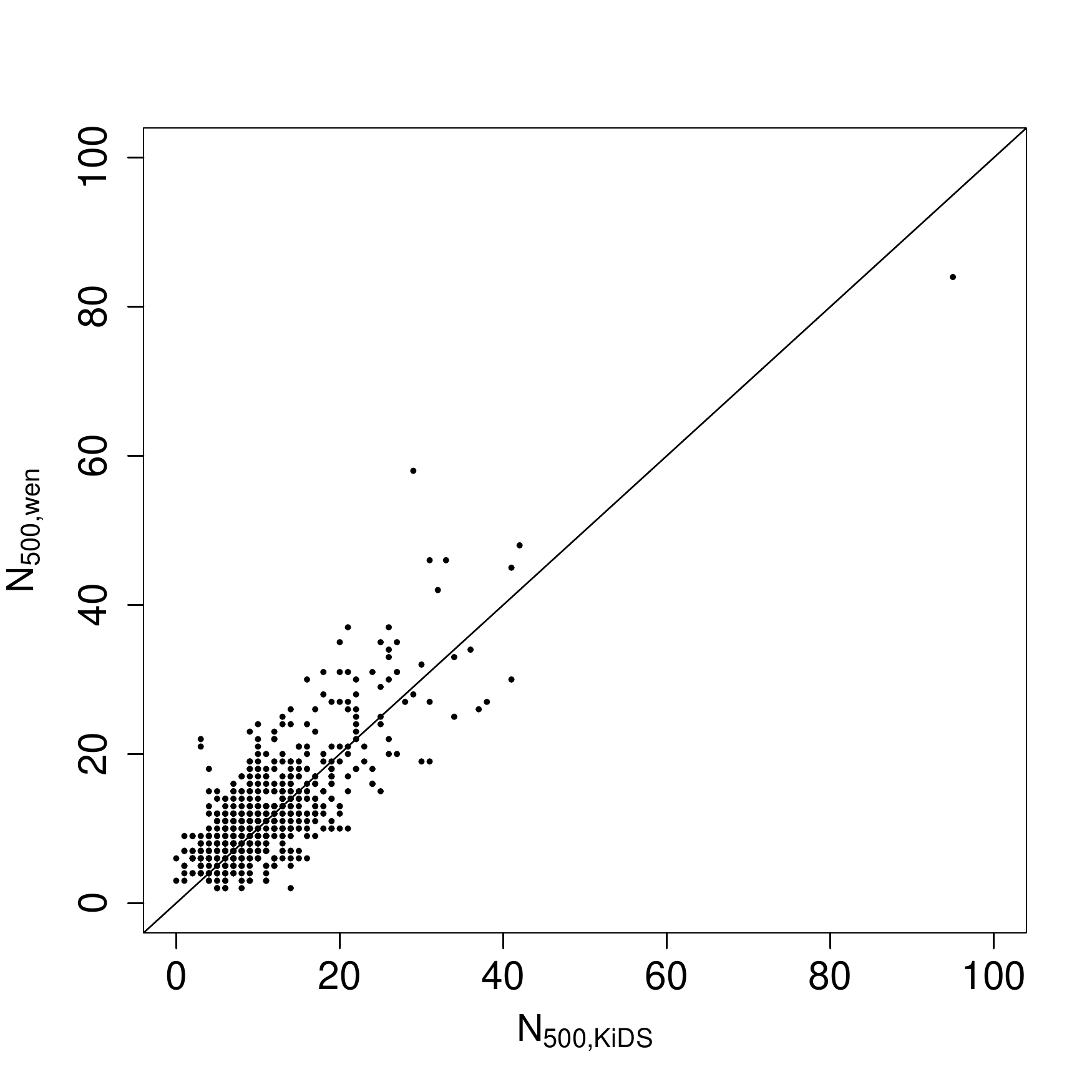}\\ 
        \caption{Comparison between the number of member galaxies found within $R=R_{500, \rm WHL15}$, in KiDS vs. WHL15. The line displays the bisector as reference.} 
        \label{fig:wen_nr}        
\end{figure}

\begin{table} 
\caption{Comparison of candidate clusters from SDSS-based catalogs in KiDS tiles and KDR2 candidate clusters. $N_{\rm SDSS}$ is the number of SDSS candidates, $N_{KiDS}$ (1858) is the number of KiDS candidates found with our method  and $N_{\rm m}$ are those matched, by position and redshift, among  our and SDSS based catalogs. The last column shows the median separation in the centers.} 
\label{tab:Nsdss} 
\begin{tabular}{ccccccc} 
\hline  \hline 
Source & $N_{\rm SDSS}$ & $N_{\rm m}$ & $N_{\rm m}/N_{\rm SDSS}$ & $N_{\rm m}/N_{\rm KiDS}$ & Sep \\ 
& & & & &  $\arcsec$ \\ 
\hline 
RM       & 293 & 226 & 77\% & 12\% & 27 \\  
AMF & 1029 & 593 & 58\% & 32\% & 87 \\ 
WHL15 &  1241 & 639 & 51\% & 34\% & 29 \\  
\hline \hline 
\end{tabular}  
 
\end{table}

\subsection*{Comparison with RM} 
\label{sec:k2rm}

The RM catalog \citep{Rykoff2014}  was derived by applying redMaPPer, a red-sequence cluster finder algorithm,  to $\sim$ 10,000 deg$^2$ in the  SDSS DR8: it consists  of approximately $ 25,000$ clusters with masses $>10^{14}$ $M_\odot$  in a redshift range $0.08 \le z \le 0.55$. The cluster catalog contains the cluster sky position, redshift, and the richness estimate ($\Lambda$) and includes only clusters with  redshift $z > 0.1$ and a richness $\Lambda > 20$, below which  the cluster completeness is shown to be lower than 50\%.  A separate catalog provides  the galaxies identified as members of each cluster (coordinates, membership probability and de-reddened magnitudes).

 A catalog of galaxies that are likely cluster red members is also available; this  enabled us to select  the RM clusters for which at least 80\% of the galaxy members are also detected  in KiDS, and reject RM clusters located close to the borders of KiDS tiles, or on masked regions. After this selection,  we obtain 293 RM clusters in our area.    
 We find that 77\% (226/293) of RM clusters are also found in KiDS. Of the 226 matched clusters, the separation is $<$ 5 arcmin for $\sim$ 99\% of them, with a median separation of $\sim$ 27  arcsec.   
  
Figure~\ref{fig:rm_zcomp} compares KiDS with RM cluster redshifts; selecting those clusters for which the redshift is $z < 0.5$, we obtain  $\sigma (\Delta z/(1+z)) = 0.02$  for both RM photometric and spectroscopic redshifts. 
  
The redshift distributions of the RM (in red) and KiDS clusters (green: all KiDS clusters) are compared in Fig.~\ref{fig:rm_redshift_all}: we detect a significantly higher number of clusters in KiDS than in RM at all redshifts. 
In order to understand the reason for this difference, we extracted a subsample of KiDS  cluster member galaxies for which the best-fit template corresponds to CWW early-type galaxies, galaxies that are located within 1 h$^{-1}$ Mpc from the cluster center, and galaxies that have an  $i-$band magnitude brighter than $m_* + 1.75$ mag \citep{Rykoff2014}.  In this way, we define a richness $n_{\rm et}$ that can be used for comparison with the richness derived in redMaPPer. We then selected those clusters  with $n_{et}>20$; the median S/N for the clusters below this limit is approximately 6. The distribution obtained after this cut is displayed in blue in Fig.~\ref{fig:rm_redshift_all}, showing that the RM and KiDS distributions are now much closer, up to $z \sim 0.3$. The number of KiDS vs. RM clusters increases at higher redshifts, as expected due to  the different depths  of the parent datasets.

\subsection*{Comparison with AMF} 
\label{sec:k2amf}

 The AMF catalog consists of  69,173 clusters in the redshift range 0.045 $\le z < 0.78$ covering an area $\sim$ 8420 deg$^2$ from SDSS DR6: for each cluster, it provides the position of the cluster center and its redshift, the richness estimate ($\Lambda_{200}$),  defined as the total luminosity in units of $L^*$ within the radius $R_{200}$. Cluster centers are defined as the position that maximizes the cluster detection probability  along a grid of resolution 1 $h^{-1}$ kpc around the initial position.  Clusters are only included in the catalog if their richness is $\Lambda_{200} > 20$, which should produce a completeness $\sim$ 85\% for clusters with $M_{200} > 10^{14}$ $h^{-1}$ $M_\odot$, based on simulations  \citep{Dong2008}. A catalog of the three brightest galaxies in the $r$ band in each cluster ($\sim$ 205,000 galaxies) is also available.    
 
1029 AMF clusters are included in the KDR2 area. Of these, 593 (58\%) are matched by KiDS clusters. The AMF radius ($R_{200}$) and richness ($\Lambda_{200}$) show a good correlation with those derived for the KiDS clusters $N_{200}$ (Pearson correlations: $r=0.6$ and $r=0.7$ respectively). The distance between the centers of the AMF clusters and  the KiDS clusters (Table \ref{tab:Nsdss}) is larger than that found for the RM. This  is due to the cluster center definition adopted in AMF, which is not related to a BCG galaxy, as in RM, WHL15, and in our case.  
 
\subsection*{Comparison with WHL15} 
\label{sec:k2WHL15} 
 
The WHL15 catalog consists of  
132,684 clusters in the redshift range $0.05 \le z \le 0.8$ from SDSS DR12,  providing the sky position of the BCG, which defines the cluster center,  and its $r-$band magnitude, the cluster redshift, the radius $R_{500}$  and the number of cluster members within this radius ($N_{500}$), and the richness estimate ($R_{L*,500}$ ).   
The scaling relations between the total $r$-band luminosity of member galaxies within 1 Mpc and the cluster radius   were established using a sub-sample of 1191 clusters with  mass derived by X--ray or Sunyaev-Zeldovich measurements. No richness cut is applied to the cluster catalog. 
 
There are 1241 WHL15 clusters in our KiDS tiles, 639 (51\%) of which are matched in KiDS; for comparison, 226 (77\%) are matched by RM clusters. Of the 602 WHL15 clusters {not} found in KiDS, 59 are matched by RM clusters. 
 
The comparison of $R_{500}$, $N_{500}$ for the clusters matched in WHL15 and KiDS gives a good correlation (Pearson correlations: $r=0.5$ and $r=0.6$ respectively). In order to compare the selection of member galaxies, we assumed the $r_{500}$ value from the WHL15 catalog  for 
each cluster, and counted the KiDS member galaxies  $r \le r_{500}$ from the BCG. As in WHL15, we only further considered  those galaxies with an absolute magnitude $M_{\rm abs}(r) + 1.16 z < -20.5$. The result is displayed in  Fig.~\ref{fig:wen_nr}. 
 
\begin{figure}[pht] 
        \includegraphics[scale=.8,viewport=150 390 470 675,clip=true]{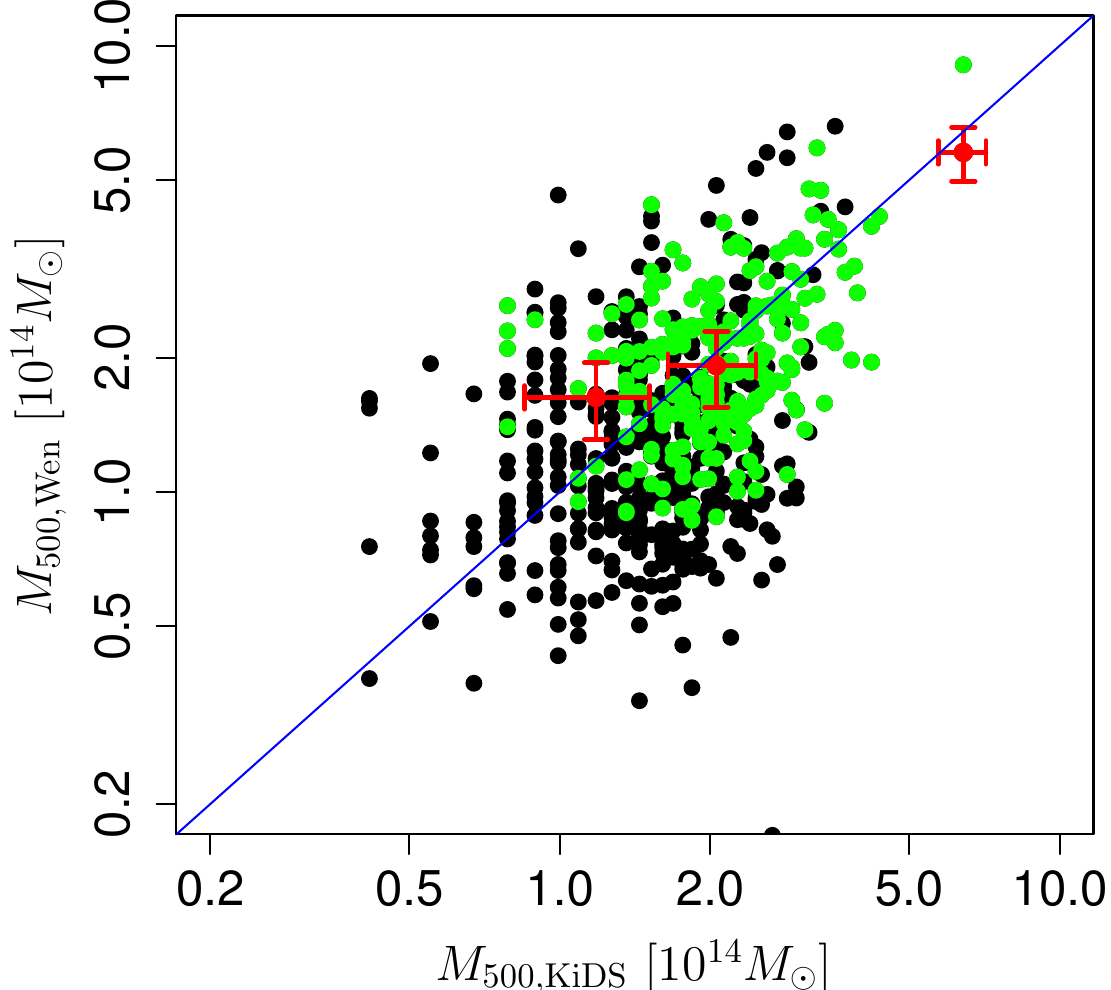}\\ 
        \caption{Comparison of KiDS and WHL15 $M_{500}$ (in units of $10^{14}$ $M_\odot$); green dots are the clusters used for the mass calibration. Red dots with error bars are the clusters found in the current KiDS area from the catalog of 1191 clusters in WHL15. The  line is the bisector.  } 
        \label{fig:wen_m500comp}        
\end{figure}

Figure~\ref{fig:wen_m500comp} compares the KiDS and WHL15 values of $M_{500}$ for the matched clusters; the red dots display $M_{500}$ for the three clusters from the WHL15 catalog of 1191 clusters with known masses. The Pearson correlation is $r=0.6$.

 
\section{Comparison with Abell, XMM, and Planck catalogs} 
\label{sec:other} 
 
One way of making a crucial assessment of the reliability of our detection algorithm and of the  cluster candidate catalog is to recover confirmed clusters included in the same area. To this end, cluster identifications were matched with the cluster catalogs of Abell \citep[ACO,][]{ACO89}, XMM Cluster Survey \citep[XCS,][]{Mehrtens2012}, and Planck-SZ \citep[PSZ1,][]{Planck2014, Planck2015}; a matching radius of 1.5 $h^{-1}$ Mpc was used here, to take into account the larger uncertainties in the centers of the X-ray and PSZ1 clusters. Given the limited size of the KiDS area analyzed here, only a few of these clusters are found. 
 
\subsection*{Abell clusters} 
 
In the KDR2 area, there are 11 Abell clusters; 9 of them are unambiguously  matched by KiDS clusters, the separation between their centers being $<$ 2\arcmin. Table \ref{TabAbells} reports  their properties \citep{ACO89, Abell58} and those derived here for the KiDS matched clusters. For the matching, we used the cluster centers listed in \citet{2dFpaper}, when available, since they are more accurate than those given in \citet{ACO89}. For two clusters (Abell 1389 and Abell 1419), the matching is more uncertain, as  the separation  is large ($\sim$ 7 \arcmin). Abell 1389 ($z = 0.08$) belongs to the Leo A supercluster \citep{1997A&AS..123..119E}, that also includes Abell 1386, which may explain the uncertainty on the cluster center.   
In the case of Abell 1419, the redshift is significantly different ($z_{\rm ACO}$ = 0.12, $z_{\rm KiDS}$ = 0.27). However, a cluster at the same position and redshift as the KiDS candidate is also found in WHL15 (WHL J115610.8-002101): this would imply that the detection of Abell 1419 is masked by a  cluster at higher redshift.   
 
The field around four  Abell clusters is displayed in Fig.~\ref{fig:ACO}.

\subsection*{XMM clusters} 
The XCS-DR1 \citep{Mehrtens2012} consists of 503 optically confirmed clusters and includes estimates of $R_{200}$ and $R_{500}$ based on X-ray data \citep{Lloyd-Davies2011}, and photometric or spectroscopic redshifts from optical identifications. Four XCS clusters fall in the KDR2 area (Table \ref{TabXMMs}): all of them are matched within the criteria defined above. For at least three of them, their redshifts (based on the SDSS: two photometric, one spectroscopic) agree well with the values found in KiDS.  No information on the accuracy of the redshift ($z = 0.15$) of XCS J1448.1-0025 \citep[from][]{2002AJ....123.1807G} is available, preventing a meaningful comparison with the KiDS redshift ($z=0.23$).

\subsection*{Planck-SZ clusters} 
There are  three Planck clusters \citep{Planck2014, Planck2015} in the KDR2 area (Table \ref{TabPlancks}); taking into account  the uncertainties on the cluster center given in the Planck catalog, the matching radius is more relaxed and therefore all Planck clusters present in our area can be considered to have a KiDS counterpart inside errors. 
The validation status parameter (column 4) provides a class of reliability for the new Planck detections. Two of our three Planck counterparts were also detected by \citet{Zwicky}, while the third is a Planck candidate of intermediate level of reliability. Also displayed in  Table~\ref{TabPlancks} are $M^{Y_Z}_{500}$, the mass derived from the SZ mass proxy, and $M_{500}$, the mass derived in Sec.~\ref{sec:results},  however, with only two clusters, no meaningful comparison can be made yet.  
 
\section{Conclusions} 
\label{sec:conclusions} 
 
In this paper, we present the methods adopted for the detection of clusters in the KiDS survey, based on an optimal filtering technique, and the initial results from  KDR2, giving a catalog of 1858 clusters with redshift $0 < z< 0.7$. The signal to noise threshold was selected by randomizing the position of galaxies and deriving the number of spurious detections as a function of S/N; we chose a threshold S/N=3.5 to minimize the number of incorrect identifications. The completeness ($\sim 85\%$) of the catalog is derived selecting clusters at low redshift from the data and simulating the effect of moving them to higher redshifts.  An estimate of the mass at $R_{500}$ and $R_{200}$ is done using the richness as a proxy; this requires calibration of  Eq.~\ref{eq_a15} for our data. The accuracy  of this calibration is still limited; we expect to improve it when more clusters with accurate masses are observed within the KiDS area. 
We present the results of the comparison between our catalog and those derived on the SDSS in the same area, showing an agreement for $> 50\%$ of the clusters.  Most of the candidate clusters present in the SDSS catalogs, but not found by us, are those with a low richness. A list of the  clusters from the Abell, XMM, and Planck-SZ surveys  included inside the KDR2 tiles is presented in Sec.~\ref{sec:other}. 
 
The results outlined in this paper are preparatory to building a statistically significant sample of clusters with richness and mass measurements. An area of 450 sq. degrees is now publicly available  (KiDS ESO-DR3); the analysis of these new data for the cluster search  is in progress, aiming to derive, for example, the mass function at different redshift bins and then to extract constraints on the main cosmological parameters.  
 
%
%

\begin{acknowledgements} 
MRa, FB, LM and MRo  acknowledge the grants ASI n.I/023/12/0 “Attivit\`a relative alla fase B2/C per la missione Euclid” and MIUR PRIN 2010-2011 “The  dark Universe and the cosmic evolution of  
baryons: from current surveys to Euclid”.  FB, LM and MRo  acknowledge the grants PRIN INAF 2012 “The  Universe in the box: multiscale simulations of cosmic structure”.  MM has been supported in part by the Transregio-Sonderforschungsbereich TR 33 of the Deutsche Forschungsgemeinschaft. MB is supported by the Netherlands Organization for Scientific Research, NWO, through grant number 614.001.451,  and through FP7 grant number 279396 from the European Research Council. 
 
We thank the referee for his/her comments, which improved the paper. 
\end{acknowledgements} 
 
\bibliography{mradovich} 
 
\newpage 
\onecolumn 
\begin{appendix} 
\section{Properties of Abell, XMM and Planck clusters} 
\label{app:A} 
 
\begin{figure*}[hb] 
        \centering 
        \includegraphics[viewport=50 1 470 395, clip=true, width=0.49\textwidth]{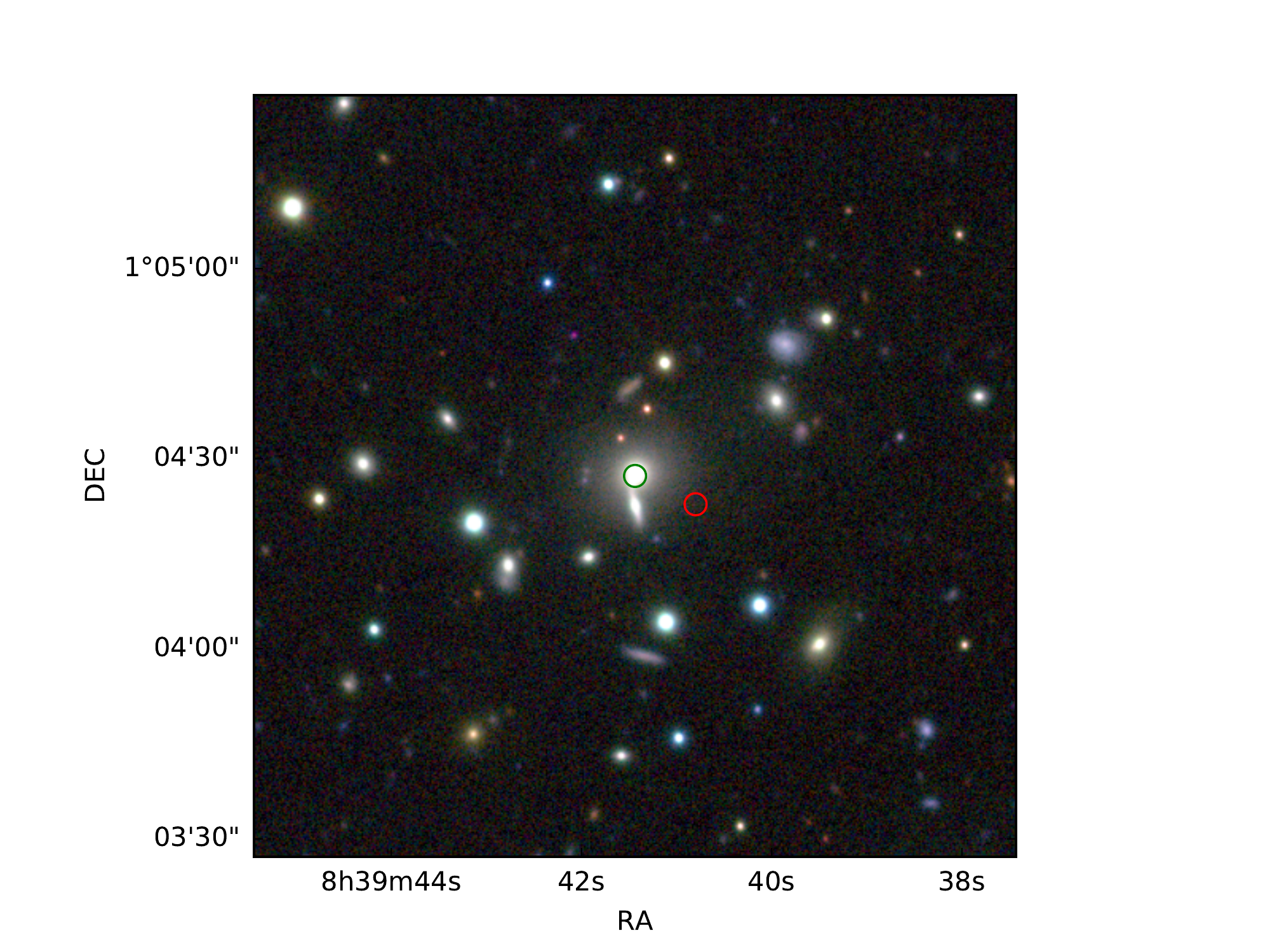} 
        \includegraphics[viewport=50 1 470 395, clip=true, width=0.49\textwidth]{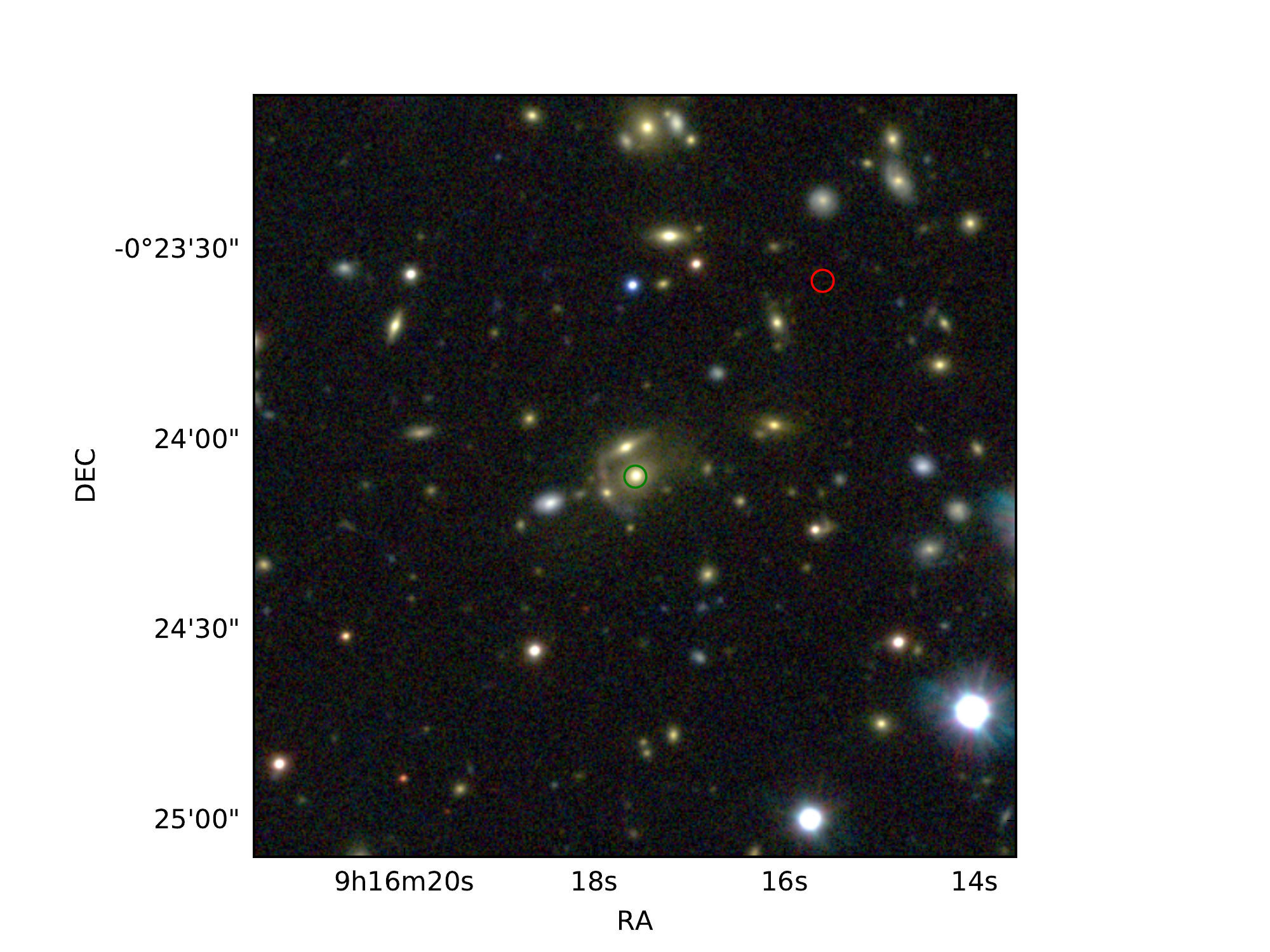} 
        \includegraphics[viewport=50 1 470 395, clip=true, width=0.49\textwidth]{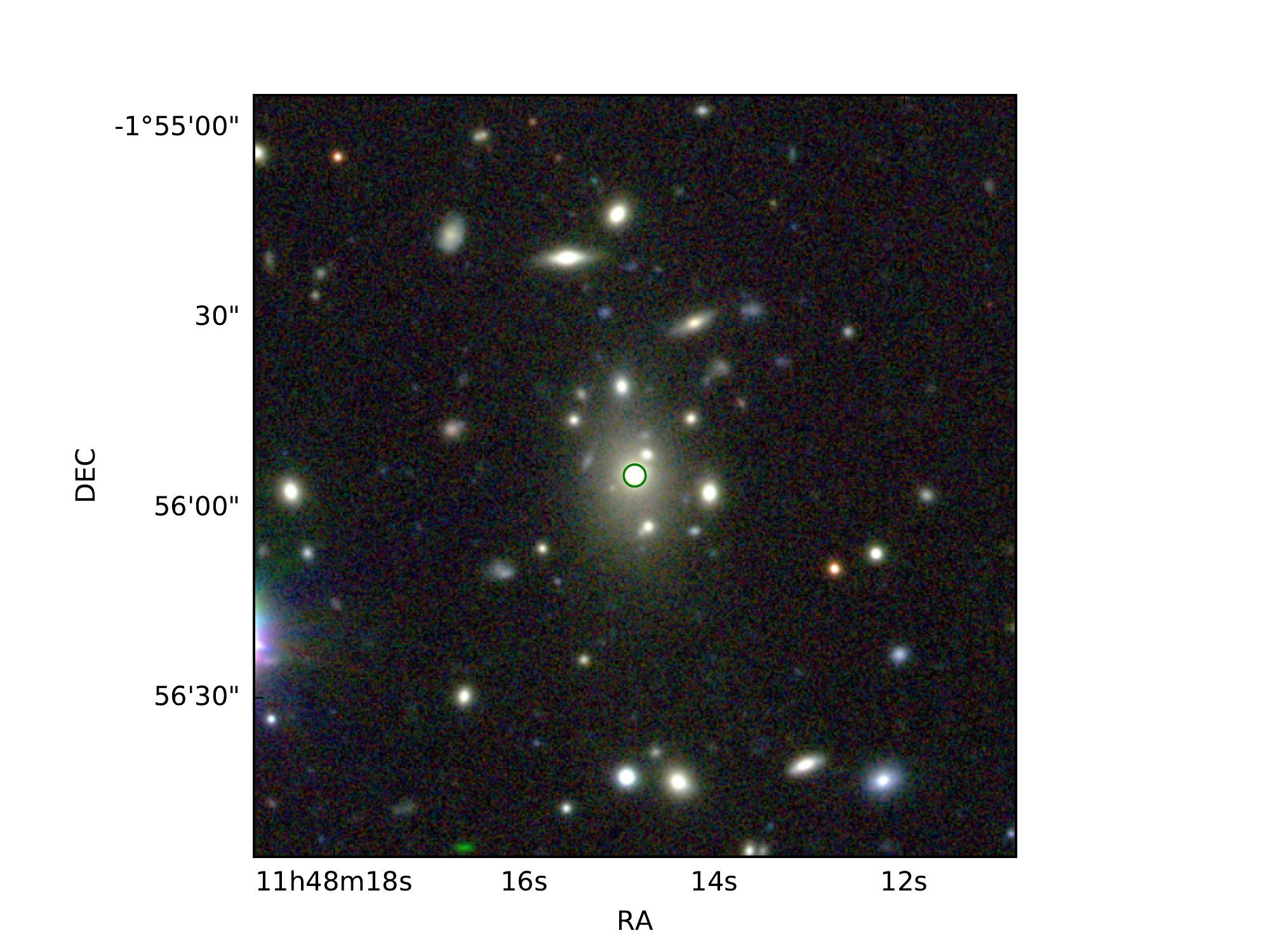} 
        \includegraphics[viewport=50 1 470 395, clip=true, width=0.49\textwidth]{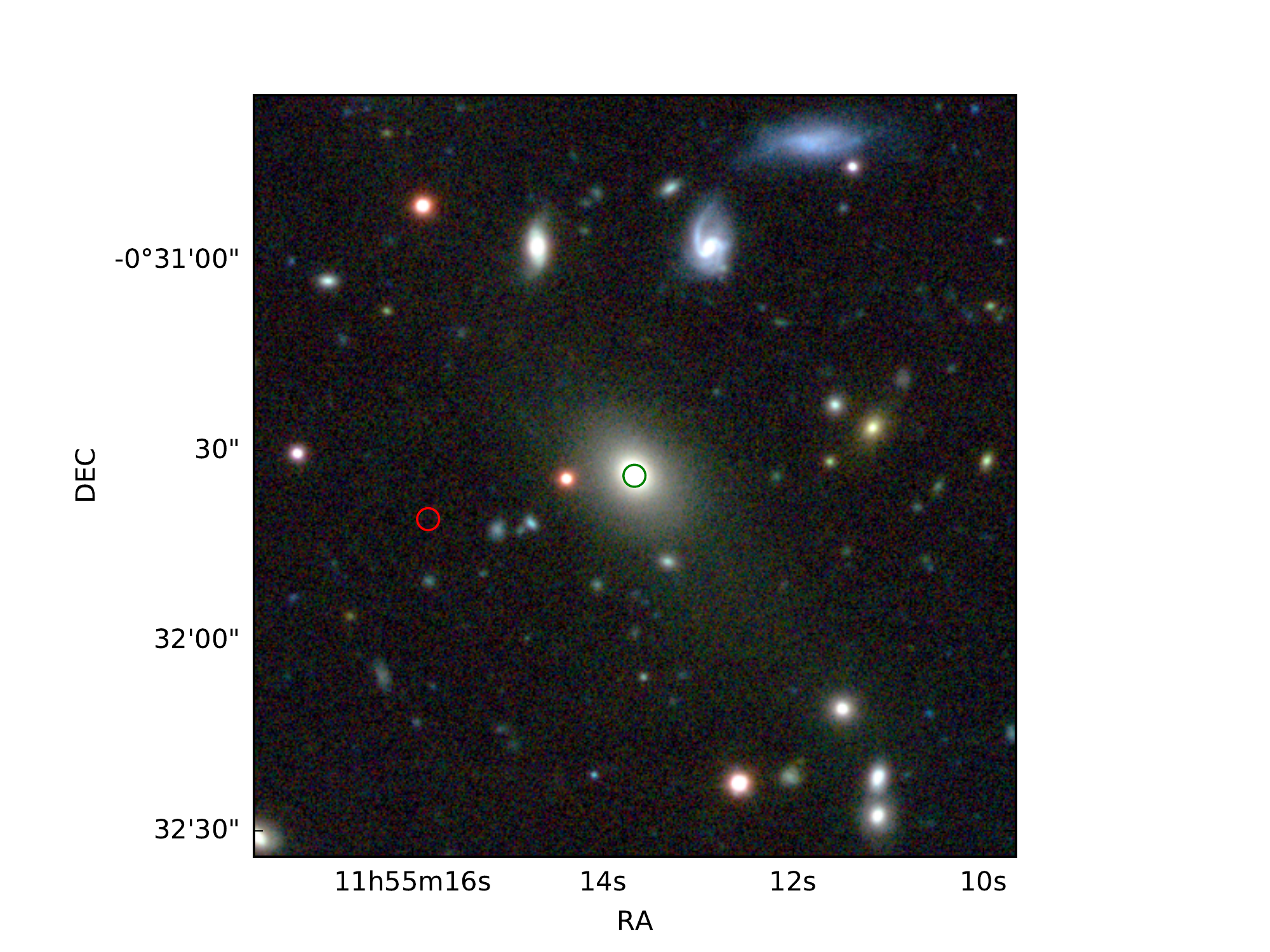}

        \caption{ KiDS $ugr$ images of the inner 2\arcmin \ regions for the Abell clusters A693, A776, A1386 and A1411 (from upper-left clockwise). Green circles show the position of the BCG; the cluster centers in the Abell catalog are displayed by the red circles, if within the region. 
        } 
        \label{fig:ACO} 
\end{figure*} 
 
\begin{table*}[hb] 
        \caption{Matching between Abell\citep{ACO89} and KiDS clusters. The last column (Sep.) gives the distance in arcsec between the centers of the Abell cluster and the KiDS cluster.} 
        \label{TabAbells} 
        \centering 
        \begin{tabular}{c c c c c c c c c c c c c c c c c} 
                \hline\hline 
                & \multicolumn{4}{c}{Abell} && \multicolumn{4}{c}{KiDS}\\ 
                \cline{2-5}  \cline{7-10} 
                 
                ID & RA$_{    }$ $^{(*)}$ & DEC$_{    }$ $^{(*)}$ & richness$^{(**)}$ & $z^{(***)}$  &&  RA&  DEC & $z$ & $S/N$ & Sep.  ($\arcsec$)\\ 
                 
                \hline 
A693  & 129.920 & 1.073 &   46 [0]  & 0.16 &&   129.923 & 1.074 & 0.17 & 7.9 & 11 \\  
A711  & 132.592 & 0.313 &   57 [1]  & 0.19 &&   132.600 & 0.260 & 0.19 & 10.6 & 192 \\  
A776  & 139.065 & -0.393 &  51      & 0.33 &&   139.073 & -0.402 & 0.37 & 16.4 & 43 \\  
A1376  & 176.529 & -1.094 &   50 [1]  & 0.12 &&   176.564 & -1.096 & 0.12 & 8.9 & 126 \\  
A1386  & 177.090 & -1.945 &   66 [1]  & 0.10 &&   177.062 & -1.932 & 0.18 & 9.4 & 112 \\  
A1411  & 178.816 & -0.528 &  69      & 0.13 &&   178.807 & -0.526 & 0.15 & 8.5 & 33 \\  
A1445  & 180.431 & -0.184 &   81 [2]  & 0.17 &&   180.432 & -0.184 & 0.19 & 13.2 & 4 \\  
A1533  & 186.140 & 0.906 &  119 [2]  & 0.23 &&   186.122 & 0.898 & 0.25 & 7.2 & 73 \\  
A1938  & 219.432 & -0.316 &  53      & 0.14 &&   219.433 & -0.316 & 0.14 & 10.1 & 2 \\  
A1389  & 177.281 & -1.367 &   40 [0]  & 0.08 &&   177.187 & -1.284 & 0.12 & 6.6 & 452 \\  
A1419  & 179.083 & -0.206 &   73 [1]  & 0.12 &&   179.041 & -0.326 & 0.27 & 12.7 & 457 \\                  
                \hline 
                \hline 
        \end{tabular} 
        \tablefoot{ 
                \tablefoottext{*}{RA and DEC for the clusters A1376, A1389, A1419, A1445 and A1938 are from \cite{2dFpaper}. The others are from \cite{ACO89}. }\\ 
                \tablefoottext{**}{Richness as defined in \cite{ACO89} (number of cluster members between $m_3$ and $m_3+2$, corrected for background) and, in square 
                        brackets, richness class from \cite{Abell58}.}\\ 
                \tablefoottext{***}{Redshift from different references, indicated in brackets; (5), (6) and (7) are photometric redshifts.}\\ 
                 
        } 
        \tablebib{ 
                (1)~\citet{1999ApJS..125...35}; 
                (2)~\citet{1995ApJS...96..343}; 
                (3)~\citet{Popesso2007}; 
                (4){Private Communication, Merchan, M. and Zandivarez, A. 
                        Unpublished Galaxy Groups catalog as described in \citet{2002MNRAS.335..216M} and sent to Dr. John Mulchaey in 2003. Received by NED in 2007}; 
                (5)~\citet{Koester2007}; 
                (6)~\citet{2010ApJS..187..272}; 
                (7)~\citet{2007ApJ...660.1176}; 
                (8) \citet{2008ApJS..176..414};. 
        } 
\end{table*}

\begin{table*} 
        \caption{Matching between XCS \citep{Mehrtens2012} and KiDS clusters.} 
        \label{TabXMMs} 
        \centering 
        \begin{tabular}{c c c c c c c c c c c c c c} 
                \hline\hline 
                &\multicolumn{5}{c}{XCS} && \multicolumn{6}{c}{KiDS}\\ 
                \cline{2-6}  \cline{8-13} 
                                 
                ID   & RA & DEC & $z^{(**)}$ & $R_{500}^{(**)}$ & $R_{200}^{(**)}$ &&  RA & DEC & $z$ & $S/N$ & 
                $R_{500}$ & $R_{200}$ & Sep.\\ 
                 &  &             & &  (kpc)          &(kpc) &   &             &      & && (kpc) & (kpc) &  ($\arcsec$) 
                \\ 
                \hline 
                 
                J0841.4+0046 & 130.352 & 0.777   & 0.41  & 565 & 857 &&   130.351 & 0.776 & 0.44 & 4.7 & 638 & 833 & 4  \\  
                J1151.5+0148 & 177.882 & 1.804   & 0.17  & 691 & 1048 &&   177.917 & 1.760 & 0.16 & 4.8 & 708 & 924 & 202  \\  
                J1225.4+0042 & 186.367 & 0.708   & 0.24  & 565 & 858 &&   186.364 & 0.710 & 0.24 & 8.7 & 828 & 1149 & 14  \\  
                J1448.1-0025 & 222.047 & -0.419   & 0.15  & - & - &&   222.133 & -0.361 & 0.23 & 3.9 & 652 & 853 & 374  \\  
                \hline 
                \hline 
        \end{tabular} 
        \tablefoot{ 
                \tablefoottext{*}{Also in \cite{2002AJ....123.1807G}}\\ 
                \tablefoottext{**}{Redshifts, $R_{200}$ and $R_{500}$ from \cite{Mehrtens2012}: the photometric redshift for J1448.1-0025 is from  \cite{2002AJ....123.1807G}.}\\ 
        } 
         
\end{table*}

\begin{table*} 
        \caption{Matching between PSZ1 \citep{Planck2015} and KiDS clusters.} 
        \label{TabPlancks} 
        \centering 
        \begin{tabular}{c c c c c cc c c c c c c c c} 
                \hline\hline 
                & \multicolumn{6}{c}{PSZ1} && \multicolumn{5}{c}{KiDS}\\ 
                                \cline{2-7}  \cline{9-13} 
                                 
                ID  & RA & DEC  & SNR  & val. & $z$ & $M^{Y_Z}_{500}$ &&   RA & DEC & $z$ & $S/N$ & $M_{500}$ & Sep.\\ 
                          & &                 & &  $^{(**)}$           & & $^{(***)}$ & && &       & & $^{(***)}$ &  ($\arcsec$)\\ 
                \hline 
                G230.73+27.70 & 135.373 & -1.658  & 5.36 & 20 & 0.29  & 5.2 $\pm$ 0.6 &&   135.393 & -1.611 & 0.33 & 13.2 & 2.9 $\pm$ 0.5 & 184 \\  
                G232.76+32.70 & 140.529 & -0.441  & 4.62 & 20 & 0.32  & 4.6 $\pm$ 0.7 &&   140.496 & -0.393 & 0.36 & 10.0 & 1.6 $\pm$ 0.4 & 210 \\  
                G286.25+62.68 & 185.293 & 0.793  & 5.52 & 2 & -  & -  &&   185.310 & 0.851 & 0.22 & 5.2 & 1.4 $\pm$ 0.3 & 219 \\          
      \hline 
                \hline 
        \end{tabular} 
        \tablefoot{ 
                \tablefoottext{*}{Also in \citet{Zwicky}}\\ 
                \tablefoottext{**}{Validation status class \citep{Planck2015}: 1 = candidate of class 1 (high reliability); 2 = candidate of class 2; 3 = candidate of class 3 (low reliability); 10 = Planck cluster confirmed by follow-up; 
                        20 = known cluster}.\\ 
                \tablefoottext{***}{$M^{Y_Z}_{500}$  and $M_{500}^{KiDS}$ are in units of $10^{14}$ $M_{\odot}$. }\\ 
                 
        } 
\end{table*} 
 
\clearpage 
\section{The KDR2 cluster catalog} 
\label{app:B} 
\begin{table}[h!] 
 
\caption{An excerpt from the cluster catalog. The full catalog is available electronically. }\label{tab:xclustcat} 
    \centering 
        \begin{small} 
    \setlength\tabcolsep{2.5pt} 
        \noindent\begin{tabular}{lccccccccccccccc} 
                \hline\hline 
                Name& ra & dec & z & snr & ra\_bcg & dec\_bcg & r\_bcg & $N_{500}$ & $r_{500}$  & $M_{500}$  & $\Delta M_{500}$  & $N_{200}$ & $r_{200}$  & $M_{200}$  & $\Delta M_{200}$  \\  
                &(deg)&(deg)&&&(deg)&(deg)&(mag)&& (Mpc) & ($10^{14} M_\odot$) & ($10^{14} M_\odot$)&&(Mpc)&($10^{14} M_\odot$)&($10^{14} M_\odot$)\\ 
                \hline 
KDR2\_J083425.9-004705 & 128.61 & -0.78 & 0.32 & 5.01 & 128.60 & -0.78 & 18.47 &   7 & 0.63 & 0.99 & 0.31 &   9 & 0.83 & 0.89 & 0.28 \\  
KDR2\_J083613.0-002823 & 129.05 & -0.47 & 0.39 & 4.24 & 129.04 & -0.47 & 19.21 &   8 & 0.63 & 1.09 & 0.33 &  13 & 0.89 & 1.20 & 0.32 \\  
KDR2\_J083628.6-003310 & 129.12 & -0.55 & 0.40 & 5.45 & 129.13 & -0.58 & 19.17 &   7 & 0.61 & 0.99 & 0.32 &  10 & 0.82 & 0.97 & 0.29 \\  
KDR2\_J083706.2-001717 & 129.28 & -0.29 & 0.42 & 5.71 & 129.28 & -0.29 & 19.15 &  13 & 0.70 & 1.53 & 0.37 &  23 & 1.03 & 1.92 & 0.40 \\  
KDR2\_J083651.8-003303 & 129.22 & -0.55 & 0.43 & 4.79 & 129.21 & -0.54 & 19.09 &   8 & 0.62 & 1.09 & 0.32 &  11 & 0.84 & 1.05 & 0.30 \\  
KDR2\_J083704.1-002657 & 129.27 & -0.45 & 0.43 & 4.46 & 129.25 & -0.47 & 18.98 &   3 & 0.50 & 0.55 & 0.24 &   7 & 0.74 & 0.72 & 0.26 \\  
KDR2\_J083541.8-001747 & 128.92 & -0.30 & 0.43 & 4.93 & 128.92 & -0.29 & 19.69 &   7 & 0.60 & 0.99 & 0.32 &   9 & 0.79 & 0.89 & 0.29 \\  
KDR2\_J083414.9-002745 & 128.56 & -0.46 & 0.44 & 6.36 & 128.56 & -0.46 & 19.24 &  21 & 0.78 & 2.13 & 0.42 &  25 & 1.04 & 2.06 & 0.40 \\  
KDR2\_J083757.4-003619 & 129.49 & -0.61 & 0.45 & 6.65 & 129.50 & -0.61 & 18.89 &   5 & 0.56 & 0.79 & 0.27 &   6 & 0.70 & 0.64 & 0.24 \\  
KDR2\_J083555.7-001250 & 128.98 & -0.21 & 0.46 & 3.84 & 128.99 & -0.21 & 19.83 &   3 & 0.49 & 0.55 & 0.24 &   5 & 0.67 & 0.55 & 0.22 \\  
KDR2\_J083620.2-001321 & 129.08 & -0.22 & 0.47 & 6.62 & 129.08 & -0.22 & 18.94 &   7 & 0.60 & 0.99 & 0.31 &  11 & 0.82 & 1.05 & 0.31 \\  
KDR2\_J083433.8-001925 & 128.64 & -0.32 & 0.52 & 7.72 & 128.64 & -0.32 & 19.52 &   6 & 0.56 & 0.89 & 0.30 &  12 & 0.83 & 1.13 & 0.31 \\  
KDR2\_J083733.4-000640 & 129.39 & -0.11 & 0.55 & 3.93 & 129.39 & -0.11 & 20.12 &   3 & 0.47 & 0.55 & 0.24 &   4 & 0.60 & 0.46 & 0.20 \\  
KDR2\_J083748.7+000428 & 129.45 & 0.07 & 0.38 & 4.23 & 129.46 & 0.06 & 19.00 &   8 & 0.64 & 1.09 & 0.33 &   8 & 0.78 & 0.81 & 0.27 \\  
KDR2\_J083422.8+000152 & 128.59 & 0.03 & 0.42 & 4.70 & 128.59 & 0.04 & 19.56 &   6 & 0.59 & 0.89 & 0.29 &   8 & 0.77 & 0.81 & 0.26 \\  
KDR2\_J083410.6+004243 & 128.54 & 0.71 & 0.42 & 6.74 & 128.54 & 0.71 & 18.51 &  11 & 0.67 & 1.36 & 0.34 &  15 & 0.91 & 1.35 & 0.34 \\  
KDR2\_J083429.3+000933 & 128.62 & 0.16 & 0.45 & 4.79 & 128.63 & 0.15 & 19.65 &   8 & 0.62 & 1.09 & 0.32 &  17 & 0.93 & 1.50 & 0.36 \\  
KDR2\_J083414.4+002256 & 128.56 & 0.38 & 0.45 & 7.31 & 128.56 & 0.38 & 19.18 &  11 & 0.67 & 1.36 & 0.35 &  20 & 0.98 & 1.72 & 0.38 \\  
KDR2\_J083538.4+003606 & 128.91 & 0.60 & 0.48 & 4.67 & 128.91 & 0.61 & 19.34 &  11 & 0.66 & 1.36 & 0.35 &  11 & 0.82 & 1.05 & 0.31 \\  
KDR2\_J083510.1+001534 & 128.79 & 0.26 & 0.59 & 4.33 & 128.79 & 0.26 & 20.24 &   1 & 0.36 & 0.26 & 0.15 &   6 & 0.66 & 0.64 & 0.24 \\  
KDR2\_J083556.2+002121 & 128.98 & 0.36 & 0.59 & 5.52 & 128.99 & 0.35 & 20.44 &   6 & 0.55 & 0.89 & 0.30 &   8 & 0.72 & 0.81 & 0.27 \\  
KDR2\_J083723.0-013713 & 129.35 & -1.62 & 0.36 & 4.06 & 129.35 & -1.61 & 18.60 &  10 & 0.67 & 1.27 & 0.34 &  17 & 0.97 & 1.50 & 0.35 \\  
KDR2\_J083743.4-014432 & 129.43 & -1.74 & 0.39 & 3.57 & 129.44 & -1.73 & 18.95 &   4 & 0.54 & 0.67 & 0.26 &   6 & 0.72 & 0.64 & 0.24 \\  
KDR2\_J083747.8-020007 & 129.45 & -2.00 & 0.40 & 5.90 & 129.44 & -2.00 & 19.96 &   5 & 0.57 & 0.79 & 0.28 &   7 & 0.75 & 0.72 & 0.26 \\  
KDR2\_J083743.2-015549 & 129.43 & -1.93 & 0.42 & 5.24 & 129.43 & -1.93 & 19.15 &   5 & 0.56 & 0.79 & 0.27 &  13 & 0.88 & 1.20 & 0.32 \\  
KDR2\_J083555.4-013719 & 128.98 & -1.62 & 0.42 & 3.67 & 128.98 & -1.62 & 19.33 &   5 & 0.56 & 0.79 & 0.28 &   8 & 0.77 & 0.81 & 0.27 \\  
KDR2\_J083709.4-012327 & 129.29 & -1.39 & 0.42 & 3.84 & 129.29 & -1.40 & 19.35 &   7 & 0.61 & 0.99 & 0.31 &   8 & 0.77 & 0.81 & 0.26 \\  
KDR2\_J083726.2-015512 & 129.36 & -1.92 & 0.46 & 5.64 & 129.36 & -1.92 & 20.12 &   4 & 0.53 & 0.67 & 0.27 &   4 & 0.63 & 0.46 & 0.20 \\  
KDR2\_J083637.7-014219 & 129.16 & -1.71 & 0.47 & 3.57 & 129.16 & -1.71 & 20.09 &   3 & 0.49 & 0.55 & 0.24 &   4 & 0.62 & 0.46 & 0.20 \\  
KDR2\_J083705.8-014949 & 129.27 & -1.83 & 0.54 & 6.92 & 129.25 & -1.83 & 20.24 &   6 & 0.56 & 0.89 & 0.29 &   7 & 0.71 & 0.72 & 0.25 \\  
         
                \hline 
 
        \end{tabular} 
    \end{small} 
    \end{table} 
     

\end{appendix} 
 
\end{document}